\documentclass[letterpaper,twocolumn,10pt]{article}
\usepackage{usenix19_v3}

\newif\ifdraft
\newif\ifcomments

\draftfalse
\commentsfalse
\usepackage{amsmath}
\usepackage{amsthm}
\usepackage{amssymb}
\usepackage{tcolorbox}
\usepackage{cite}
\usepackage[shortlabels,inline]{enumitem}
\usepackage{xspace}
\usepackage{upquote}
\usepackage{graphicx}
\usepackage{mdframed}
\usepackage{algorithmicx}
\usepackage{algpseudocode}
\usepackage{algorithm}
\usepackage{xcolor}
\usepackage{multirow}
\usepackage{listings}
\usepackage{courier}
\usepackage{bbding}
\usepackage{tikz}
\usepackage{hanging}
\usepackage[hyphens]{url}
\usepackage{microtype}
\usepackage[binary-units]{siunitx}
 \usepackage[colorlinks=true,
             linkcolor=black,
             urlcolor=blue,
             citecolor=blue,
             breaklinks, 
             pdfborder={0 0 0}]{hyperref}           
\usepackage[aboveskip=5pt, font={small}, labelfont=bf]{caption}
\usepackage{subcaption}
\usepackage[T1]{fontenc}
\usepackage[normalem]{ulem} %
\usepackage{comment}
\usepackage[noabbrev,capitalize,nameinlink]{cleveref}

\ifcomments
\usepackage[textsize=small,color=lightgray]{todonotes}
\else
\usepackage[disable]{todonotes}
\fi

\ifdraft
    \usepackage{fancyhdr, datetime}
\fi

\usepackage[small,compact]{titlesec}
\titleformat{\subsubsection}{\normalfont\em\bf}{\thesubsubsection}{1em}{\em}
\titlespacing*{\section}{0pt}{7pt}{3pt}
\titlespacing*{\subsection}{0pt}{4pt}{2pt}
\titlespacing*{\subsubsection}{0pt}{2pt}{0pt}

\makeatletter
\g@addto@macro\normalsize{%
\setlength{\abovedisplayskip}{2pt}
\setlength{\belowdisplayskip}{2pt}
\setlength{\abovedisplayshortskip}{0pt plus 2pt}
\setlength{\belowdisplayshortskip}{0pt plus 2pt}
}
\makeatother

\newcommand{\parhead}[1]{\noindent\textbf{#1.}\enskip}
\newcommand{\parheadskip}[1]{\smallskip\noindent\textbf{#1.}\enskip}
\renewcommand{\paragraph}[1]{\parheadskip{#1}}

\newcommand{\code}[1]{\texttt{#1}\xspace}

\definecolor{FGreen}{cmyk}{0.9,0.2,0.5,0.3}

\definecolor{mygreen}{RGB}{34,139,34}	

\makeatletter
\newcommand*{\ie}{%
    \@ifnextchar{,}%
        {\textit{i.e.}}%
        {\textit{i.e.},\@\xspace}%
}
\makeatother
\makeatletter
\newcommand*{\eg}{%
    \@ifnextchar{,}%
        {\textit{e.g.}}%
        {\textit{e.g.},\@\xspace}%
}
\makeatother
\makeatletter
\newcommand*{\etc}{%
    \@ifnextchar{.}%
        {etc}%
        {etc.\@\xspace}%
}
\makeatother
\makeatletter
\newcommand*{\etal}{%
    \@ifnextchar{.}%
        {\textit{et al}}%
        {\textit{et al.}\@\xspace}%
}
\makeatother

\crefformat{section}{\S#2{\color{blue}#1}#3}
\crefformat{subsection}{\S#2{\color{blue}#1}#3}
\crefformat{subsubsection}{\S#2{\color{blue}#1}#3}
\crefname{appsec}{Appendix}{Appendices}
\crefdefaultlabelformat{#2{\color{blue}#1}#3}

\def\approx{$\mathtt{\sim}$} %
\newcommand*\bubble[1]{\tikz[baseline=(char.base)]{
            \node[shape=circle,scale=0.8,draw,inner sep=2pt,fill=black,
            text=white] (char)
            {#1};}}

\makeatletter
\def\thickhline{%
  \noalign{\ifnum0=`}\fi\hrule \@height \thickarrayrulewidth \futurelet
   \reserved@a\@xthickhline}
\def\@xthickhline{\ifx\reserved@a\thickhline
               \vskip\doublerulesep
               \vskip-\thickarrayrulewidth
             \fi
      \ifnum0=`{\fi}}
\makeatother
\newlength{\thickarrayrulewidth}
\newenvironment{compactitemize}{ 
    \begin{itemize}[nolistsep,leftmargin=*]
        \setlength{\itemsep}{0.5pt}
        \setlength{\parskip}{0.9pt}
        \setlength{\parsep}{0.1pt}     
    }{\end{itemize}}

\newenvironment{mydescription}{ 
        \begin{description}[nolistsep]
            \setlength{\itemsep}{0.5pt}
            \setlength{\parskip}{0.9pt}
            \setlength{\parsep}{0.1pt}     
    }{\end{description}}

\newenvironment{compactenumerate}{ 
    \begin{enumerate}[nolistsep,leftmargin=*]
        \setlength{\itemsep}{0.5pt}
        \setlength{\parskip}{0.9pt}
        \setlength{\parsep}{0.1pt}     
    }{\end{enumerate}}

\newenvironment{compacttabbing}
    {\setlength{\topsep}{1pt}%
        \setlength{\partopsep}{1pt}%
    \tabbing}
    {\endtabbing}

\newtheorem{theorem}{Theorem}
\newtheorem{definition}{Definition}

\newcommand{\twopsi}{2\mbox{-}\mathsf{SI}}
\newcommand{\mpsi}{m\mbox{-}\mathsf{SI}}
\newcommand{\twopsu}{2\mbox{-}\mathsf{SU}}
\newcommand{\mpsu}{m\mbox{-}\mathsf{SU}}

\newcommand{\msort}{m\mbox{-}\mathsf{Sort}}
\newcommand{\merge}{\mathsf{Merge}}
\newcommand{\mono}{\mathsf{Mono}}
\newcommand{\twopsitagged}{\twopsi\mathsf{*}}

\newcommand{\ver}{\mathsf{Ver}}
\newcommand{\dedup}{\mathsf{Dedup}}
\newcommand{\filter}{\mathsf{Filter}}

\usepackage{relsize}

\newcommand{\Select}{\mathlarger{{\sigma}}}

\newcommand{\Sort}{\mathlarger{{\tau}}}
\newcommand{\Groupby}{\mathlarger{{\gamma}}}
\newcommand{\Agg}{\Sigma}
\newcommand{\Union}{\mathlarger{{\cup}}}

\let\tmpJoin\Join
\renewcommand{\Join}{\mathlarger{{\tmpJoin}}}

\newcommand{\fromparty}[1]{\textcolor{mygreen}{#1}}
\newcommand{\fp}[1]{\fromparty{#1}}

\newcommand{\Reln}[2]{#1\fp{|#2}}

\newcommand{\partyset}{{\mathcal{P}}}

\newcommand{\garbled}{{\sf G}}
\newcommand{\fgarble}{{\sf Garble}\xspace}
\newcommand{\fwrkgarble}{{\sf WRK\cdot Garble}\xspace}
\newcommand{\feval}{{\sf Eval}\xspace}
\newcommand{\fwrkeval}{{\sf WRK\cdot Eval}\xspace}

\newcommand{\fmpctree}{\mathcal{F}_{\sf MPC\cdot tree}}

\newcommand{\fsolder}{\mathcal{F}_{\sf Solder}\xspace}

\newcommand{\pmpctree}{\Pi_{\sf MPC\cdot tree}}

\newcommand{\psolder}{\Pi_{\sf Solder}\xspace}

\newcommand{\adv}{\mathcal{A}}
\newcommand{\simul}{\mathcal{S}}

\newcommand{\authshare}[1]{\langle #1  \rangle}

\newcommand{\parties}{{\sf parties}}
\newcommand{\inp}{\mathcal{I}}
\newcommand{\outp}{\mathcal{O}}

\newtheorem{myfunctionality}{FUNCTIONALITY}{\bf}{}
{\bf}{}
\newtheorem{myprotocol}{PROTOCOL}{\bf}{}
\crefname{myprotocol}{Protocol}{Protocols}
\crefname{myfunctionality}{Functionality}{Functionalities}
\crefname{myprocedure}{Procedure}{Procedures}

\newcommand{\eat}[1]{\ignorespaces}

\usepackage{letltxmacro}
\LetLtxMacro{\todonote}{\todo}

\renewcommand{\todo}[2][]
{\todonote[color=red!40,size=\small,caption={#2}, #1]
{#2}}

\newcommand{\maxperf}{$145\times$}

\ifcomments
\paperwidth=\dimexpr \paperwidth + 10cm\relax
\oddsidemargin=\dimexpr\oddsidemargin + 5cm\relax
\evensidemargin=\dimexpr\evensidemargin + 5cm\relax
\marginparwidth=\dimexpr \marginparwidth + 5cm\relax
\fi

\newcommand{\sys}{Senate\xspace}
\tcbset{boxrule=0.1mm}

\begin{document}

\date{}

\twocolumn[
\bigskip
\centerline{\Large \bf \sys: A Maliciously-Secure MPC Platform for Collaborative Analytics}
\bigskip
\centerline{\large Rishabh Poddar \quad Sukrit Kalra \quad Avishay Yanai$^*$ \quad Ryan Deng}
\smallskip
\centerline{\large Raluca Ada Popa \quad Joseph M. Hellerstein}
\bigskip
\centerline{\large \it UC Berkeley \quad $^*$VMware Research}
\bigskip \smallskip
]

\begin{abstract}

Many organizations stand to benefit from pooling their data together in order to draw mutually beneficial insights---\eg for fraud detection across banks, better medical studies across hospitals, \etc.
However, such organizations are often prevented from sharing their data with each other by privacy concerns, regulatory hurdles, or business competition.

We present \sys, a system that allows multiple parties to collaboratively run analytical SQL queries without revealing their individual data to each other. %
Unlike prior works on secure multi-party computation (MPC) that assume that all parties are semi-honest, \sys protects the data even in the presence of malicious adversaries.
At the heart of \sys lies a new {\em MPC decomposition protocol}  
that decomposes the cryptographic MPC computation into smaller units, some of which can be executed by {\em subsets} of parties and in {\em parallel}, while preserving its security guarantees.
\sys then provides a {\em new query planning algorithm} that decomposes and plans the cryptographic computation effectively, achieving a  performance of up to \maxperf{} faster than  the state-of-the-art.
\end{abstract}

\pagestyle{plain}

\section{Introduction}\label{s:intro}

A large number of services today collect valuable sensitive user data. %
These services could benefit from pooling their data together and jointly performing query analytics on the aggregate data. 
For instance, such collaboration can enable 
better medical studies
\cite{SMCQL:Bater:VLDB,GenomeAssociation:Kamm};
identification of criminal activities (\eg fraud)~\cite{Frauddetection:Sangers:2018};
more robust financial services~\cite{Frauddetection:Sangers:2018, Riskanalytics:Bisias, Riskexposures:Abbe, DStress:Papadimitriou};
and more relevant online advertising~\cite{PSIsum:Google:2017}.
However, many of these institutions cannot share their data with each other due to privacy concerns, %
regulations, 
or business competition.

Secure multi-party computation~\cite{Yao:82:Millionaires, GMW87, BGW88} (MPC) promises to enable such scenarios 
by allowing $m$ parties, each having secret data  $d_i$, to compute a function $f$ on their aggregate data, and to share the result $f(d_1, \dots, d_m)$ amongst themselves, {\em without} learning each other's data beyond what the function's result reveals. At a high level, MPC protocols work by having each party encrypt its data, and then perform joint computations on encrypted data leading to the desired result.

Despite the pervasiveness of data analytics workloads, there are very few works that consider secure collaborative analytics. 
While closely related works such as SMCQL~\cite{SMCQL:Bater:VLDB} and Conclave~\cite{Conclave:Volgushev:Eurosys}
make useful first steps in the direction of secure collaborative analytics, their main limitation is their weak security guarantee: {\em semi-honest security}. Namely, these works assume that each party, even if  compromised, follows the protocol faithfully.  If any party deviates from the protocol, it can, in principle, extract information about the sensitive data of other parties.
This is an unrealistic assumption in many scenarios for two reasons. 
First, each party running the protocol at their site has full control 
over what they are actually running.
For example, it requires a bank to place the confidentiality of its sensitive 
business data in the hands of its competitors.
If the competitors secretly deviate 
from the protocol, they could learn information about the bank's data without its knowledge.
Second, in many real-world attacks~\cite{SHELLSHOCK:NYT}, attackers are able to install software on the server or obtain control
of a server~\cite{DIRTYSOCK:UBUNTU}, thus allowing them to  alter the server's behavior.

\subsection{\sys overview}\label{s:overview}

We present \sys, a platform for secure collaborative analytics with the strong guarantee of {\em malicious security}. 
In \sys, even if $m-1$ out of $m$ parties fully misbehave and collude, an honest party is guaranteed that nothing leaks about their data other than the result of the agreed upon query. Our techniques come from a synergy of new cryptographic design and insights %
in query rewriting and planning.  
A high level overview of \sys's workflow (as shown in \cref{fig:workflow})
is as follows:

\begin{figure}
\centering
\includegraphics[width=\columnwidth]{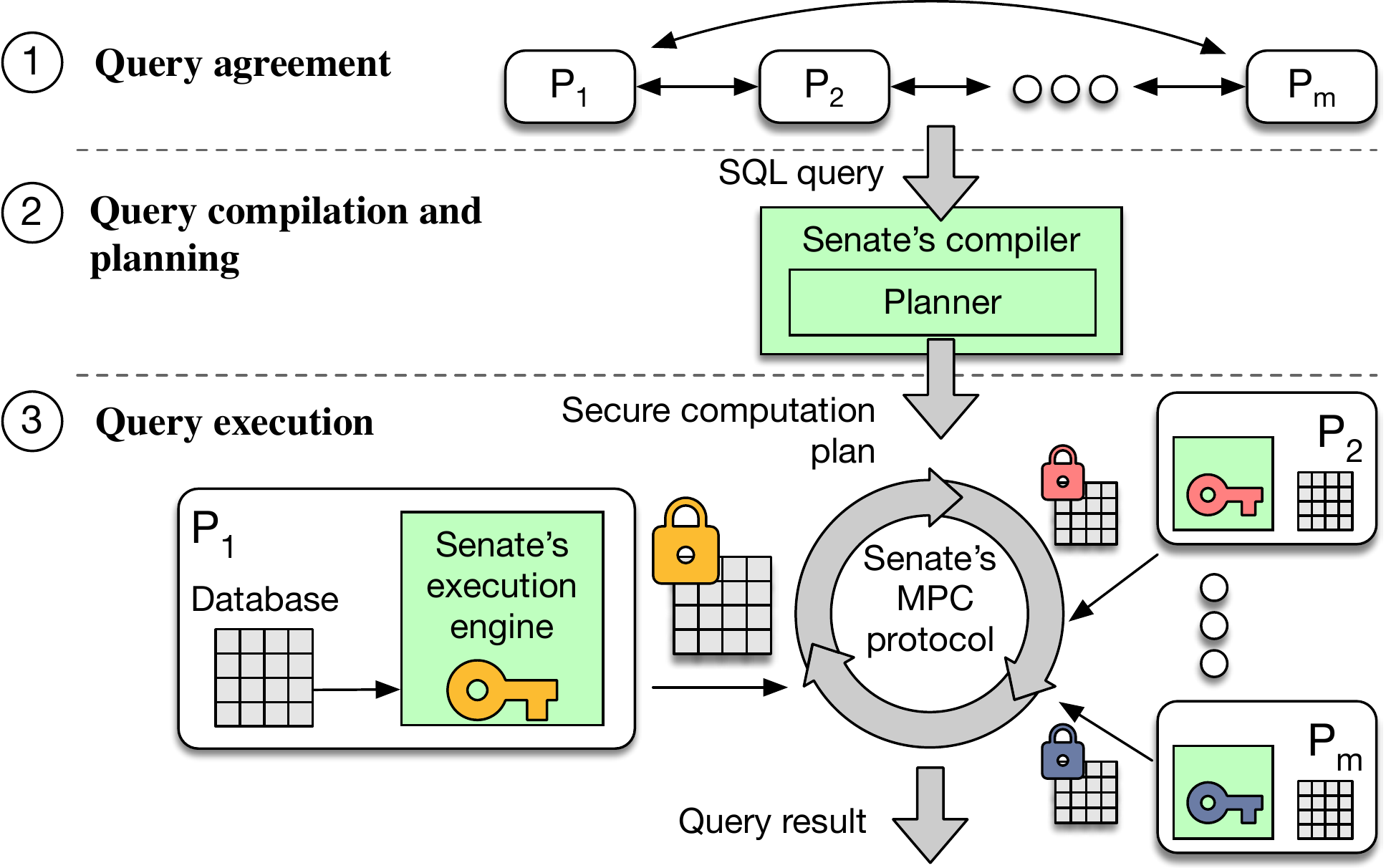}
\caption{Overview of \sys's workflow.}
\label{fig:workflow}
\end{figure}

\paragraph{Agreement stage} The $m$ parties agree on a shared schema for their data, and on a query for which they are willing to share the computation result. This happens before invoking \sys and may involve humans. 

\paragraph{Compilation and planning stage} 
\sys's compiler takes the query and certain size information (described in \cref{s:api}) as input and outputs a cryptographic execution plan.
It  runs at each party, deterministically producing the same plan.
In particular, the compiler employs our {\em consistent and verifiable query splitting} technique in order to minimize the amount of joint computation performed by the parties.
Then, the compiler plans the execution of the joint computation using our \emph{circuit decomposition} technique, which can produce a significantly more efficient execution plan.

\paragraph{Execution stage} An execution engine at each party runs the cryptographic execution plan by coordinating with the other parties and routing encrypted intermediate outputs based on the plan. 
This is done using our efficient \emph{MPC decomposition protocol}, which outputs the query result to all the parties.

\subsection{\sys's techniques}

\begin{figure}
    \centering
    \includegraphics[width=\columnwidth]{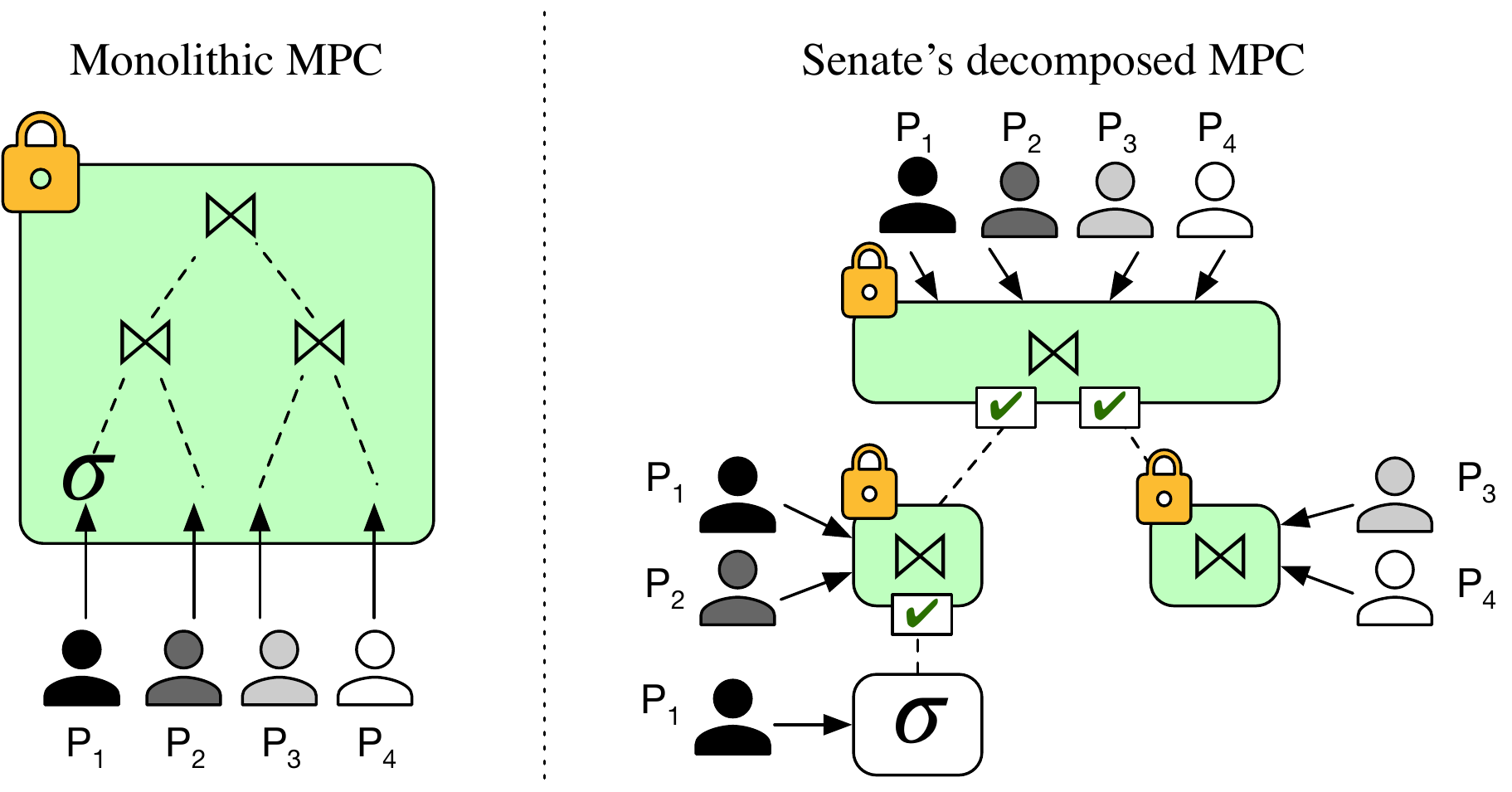}
    \caption{Query execution in the baseline (monolithic MPC) vs.\ \sys (decomposed MPC). $\Select$ represent a filtering operation, and $\Join$ is a join. 
    Green boxes with locks denote MPC operations; white boxes denote plaintext computation.
    \textcolor{mygreen}{\checkmark} represents additional verification operations added by \sys.}
    \label{fig:overview}
\end{figure}

Designing a maliciously-secure collaborative analytics system is challenging due to the significant overheads of such strong security.
Consider simply using a state-of-the-art $m$-party maliciously-secure MPC tool such as AGMPC~\cite{AGMPC:Git} which implements the protocol of Wang \etal~\cite{WRK17:Global}; we refer to this as the {\em baseline}.
When executing a SQL query with this baseline, the query gets transformed into a single, large 
Boolean circuit (\ie a circuit of AND, OR, XOR gates) taking as input the data of the $m$ parties.
The challenge then is that  the $m$ parties need to execute a {\em monolithic} cryptographic computation {\em together} to evaluate this circuit.  %

\parheadskip{Minimizing joint computation}
Prior work~\cite{SMCQL:Bater:VLDB, Conclave:Volgushev:Eurosys} in the semi-honest setting shows that one can significantly improve performance by splitting a query into local computation (the part of the query that touches only one party's data) and the rest of the computation.
The former can be executed locally at the party on plaintext, and the latter in MPC;
\eg if a query filters by ``\code{disease = flu}'', the parties need to input only the records matching the filter into MPC as opposed to the entire dataset. 
In the semi-honest setting, the parties are trusted to perform such local computation faithfully.
Unfortunately, this technique no longer works with malicious parties because a malicious party $M$ can perform the local computation: 
\begin{compactitemize}
\item  {\em incorrectly}. For example, $M$ can input  records with ``\code{disease = HIV}'' into MPC.  
    This %
    can reveal information about another party's ``\code{HIV}'' records, \eg via a later join operation, when this party might have expected the join to occur only over rows with the value ``\code{flu}''. %
    \item {\em inconsistently}. For example, if one part of a query selects patients with ``\code{age = 25}'' and another with ``\code{age $\in [20,30]$}'', the first filter's outputs should be included within the second's. 
    However, $M$ might provide inconsistent sets of records as the outputs of the two filters.
\end{compactitemize}

\sys's {\em verifiable and consistent query splitting} technique allows \sys to take advantage of local computation via a different criteria than in the semi-honest case. Given a query, \sys's compiler splits the query into a special type of local computation---one that does not introduce inconsistencies---and a joint computation, which it annotates with verification of the local computation, in such a way that the verification is faster to execute than the actual computation. 
\noindent
For example, \cref{fig:overview} shows a $4$-party query in which party $P_1$'s inputs are first filtered (denoted $\Select$). Unlike the baseline execution, \sys enables $P_1$ to evaluate the filter locally on plaintext, and the secure computation proceeds from there on the smaller filtered results; these results  are then jointly verified.

\parheadskip{Decomposing MPC}
In order to decompose the joint computation (instead of evaluating a single, large circuit using MPC) one needs to open up the cryptographic black box. Consider a 4-way join operation ($\Join$) among tables of $4$ parties, as shown in \cref{fig:overview}. With the baseline, all 4 parties have to execute the whole circuit.
However, if privacy were not a concern, $P_1$ and $P_2$ could join their tables without involving the other parties, $P_3$ and $P_4$ do the same \emph{in parallel}, and then everyone performs the final join on the smaller intermediate results. 
This is not possible with existing state-of-the-art protocols for MPC, which execute the computation in a \emph{monolithic} fashion.

To enable such decomposition, we design a new cryptographic protocol we call {\em secure MPC decomposition} (\cref{s:protocol}), which may be of broader interest beyond \sys.
In the example above, our protocol enables parties $P_1$ and $P_2$ to evaluate their join obtaining an encrypted intermediate output, and then to {\em securely  reshare} this output with parties $P_3$ and $P_4$ as they all complete the final  join. The decomposed circuits include verifications of prior steps needed for malicious security. 
We also develop more efficient Boolean circuits for expressing common SQL operators such as  joins, aggregates and sorting (\cref{s:sql}), using a small set of Boolean circuit primitives which we call $\mpsi$, $\mpsu$ and $\msort$ (\cref{s:primitives}).

\parheadskip{Efficiently planning query execution}
Finally, we develop a new {\em query planner}, which leverages \sys's MPC decomposition protocol (\cref{s:planning:algorithm}). Unsurprisingly, the circuit representation of a complex query can be decomposed in many different ways. However, the rules governing the cost of each execution plan differ significantly from regular computation. Hence, we develop a {\em cost model} for our protocol which estimates the cost given a circuit configuration (\cref{s:planning:costmodel}). 
\sys's query planner selects the most efficient plan based on the cost model.

\subsection{Evaluation summary}
We implemented \sys and evaluate it in~\cref{s:evaluation}. 
Our decomposition and planning mechanisms result in a performance improvement of up to $10\times$ compared to the monolithic circuit baseline, with up to $11\times$ less resource consumption (memory / network communication), on a set of representative queries.
\sys's query splitting technique for local computation can \emph{further} increase performance by as much as $10\times$, bringing the net improvement to up to $100\times$.
Furthermore, to stress test \sys on more complex query structures, we also evaluate its performance on the TPC-H analytics benchmark~\cite{TPCH:Web}; we find that \sys's improvements range from $3\times$ to $145\times$.

Though MPC protocols have improved steadily, they still have notable overhead. 
Given that such collaborative analytics do not have to run in real time, we believe that \sys can already be used for simpler  workloads and / or relatively small databases, but is not yet ready for  big data analytics.
However, %
we expect faster MPC protocols to  continue to appear. %
The systems techniques in \sys will apply independently of the protocol, and the cryptographic decomposition will likely have a similar counterpart.

\section{\sys's API and example queries}\label{s:api}

\sys exposes a SQL interface to the parties. %
To reason about which party supplies which table in a collaborative setting,  we \textcolor{mygreen}{augment} the query language with the simple notation 
$\Reln{R}{P}$ to
 indicate that table $R$ comes from party $P$.
 Hence, $\Reln{R}{P_1}$ $\Union$  $\Reln{R}{P_2}$ indicates that  each party holds a {\em horizontal} partition of table $R$. 
 One can obtain a {\em vertical} partitioning, for example, by joining two tables from different parties 
 $\Reln{R_1}{P_1}$  and $\Reln{R_2}{P_2}$. 
 Here, we use the $\Union$ operator to denote a simple concatenation of the tables, instead of a set union (which removes duplicates). 

   In principle, \sys can support arbitrary queries because it builds on a generic MPC tool. %
The performance improvement of our techniques, though, is more relevant to joins, aggregates, and filters.
We now give three use cases and queries, each from a different domain, which we use as running examples.

\noindent {\bf Query 1. Medical study} \cite{SMCQL:Bater:VLDB}. 
Clostridium difficile (cdiff) is an infection that is often antibiotic-resistant.
As part of a clinical research study, medical institutions $P_1\ldots P_m$ wish to collectively compute the most common diseases contracted by patients with cdiff. %
However, they cannot share their databases with each other to run this query due to privacy regulations. 
\begin{compacttabbing}
    \code{SELECT} \=  \code{diag, COUNT(*) AS cnt} \\
    \code{FROM} $\Reln{\code{diagnoses}}{P_1}$ ${\Union \dots \Union}$ $\Reln{\code{diagnoses}}{P_m}$ \\
    \> \code{WHERE has\_cdiff  = `True'} \\
    \> \code{GROUP BY diag}
    \code{ORDER BY cnt}
    \code{LIMIT 10;}
\end{compacttabbing}
\noindent {\bf Query 2. Prevent password reuse}\cite{PASSREUSE:WR19:NDSS}.
Many users unfortunately  reuse passwords across different sites.
If one of these sites is hacked, the attacker could compromise the account of these users at other sites. As studied in~\cite{PASSREUSE:WR19:NDSS}, sites wish to identify which users reuse passwords across the sites, and can arrange for the salted hashes of the passwords to match if the underlying passwords are the same (and thus be compared to identify reuse using the query below). However, these sites do not wish to share what other users they have or the hashed passwords of these other users (because they can be reversed).
\begin{compacttabbing}
    \code{SELECT} \= \code{user\_id}\\
    \code{FROM} $\Reln{\code{passwords}}{P_1}$  $\Union \dots \Union$  $\Reln{\code{passwords}}{P_m}$ \\ 
    \> \code{GROUP BY} \code{CONCAT(user\_id, password)} \\
    \> \code{HAVING COUNT(*) > 1;}
\end{compacttabbing}
\noindent {\bf Query 3.  Credit scoring} agencies do not want to share their databases with each other~\cite{Conclave:Volgushev:Eurosys} due to business competition, yet they want to identify records where they have a significant discrepancy in a particular financial year. For example, an individual could have a low score with one agency, but a higher score with another; the individual could take advantage of the higher score to obtain a loan they are not entitled to.
\begin{compacttabbing}
    \code{SELECT} \= \code{c1.ssn}\\
    \code{FROM} $\Reln{\code{credit\_scores}}{P_1}$ \code{AS c1} \\ 
    ${\dots}$\\
    \code{JOIN} $\Reln{\code{credit\_scores}}{P_m}$ \code{AS cm ON c1.ssn = cm.ssn} \\
    \code{WHERE} \= \code{GREATEST(c1.credit, $\dots$, cm.credit) -} \\
    \quad \code{LEAST(c1.credit, $\dots$, cm.credit) > threshold} \\
    \quad \code{AND c1.year = 2019 $\dots$ AND cm.year = 2019;}
\end{compacttabbing}

\subsection{Sizing information}
\label{s:api:sizing}

Given a query, \sys's compiler first splits the query into local and joint computation.
Each party then specifies to the compiler an upper bound on the number of records it will provide as input to the joint computation, following which the compiler maps the joint computation to circuits.
These upper bounds are useful because we do not want to leak the size of the parties' inputs, but also want to improve performance by not defaulting to the worst case, \eg the maximum number of rows in each table.
For example, for Query 1, \sys transforms the query so that the parties group their records locally by the column \code{diag} and compute local counts per group. %
In this case, \sys asks for %
the upper bound on the number of diagnoses per party. %
In many cases, deducing such upper bounds is not necessarily hard: \eg it is simple for Query 1 because there is a fixed number of known diseases~\cite{cdcdiseases}.  
Further, meaningful upper bounds significantly improve performance.

\section{Threat model and security guarantees}
\label{s:threatmodel}

\sys adopts a strong threat model in which a malicious adversary can corrupt $m-1$ out of $m$ parties.
The corrupted parties may arbitrarily deviate from the protocol and collude with each other.
As long as one party is honest, the only information the compromised parties learn about the honest party is the final global query result (in addition to the upper bounds on data size provided to the compiler by the parties, and the query itself).

More formally, we define an ideal functionality $\fmpctree$ (\cref{func:mpctree}, \cref{s:protocol:mpctree}) for securely executing functions represented as a tree of circuits, while placing some restrictions on the structure of the tree.
We then develop a protocol that realizes this functionality and prove the security of our protocol (per \cref{th:security:mpc}, \cref{s:protocol:mpctree}) 
according to the  definition of security for (standalone) maliciously secure MPC~\cite{Goldreich2004}, as captured formally by the following definition:

\begin{definition}
    Let $\mathcal{F}$ be an $m$-party functionality, and let $\Pi$ be an $m$-party protocol that computes $\mathcal{F}$.
    Protocol $\Pi$ is said to {\em \sf securely compute} $\mathcal{F}$ in the presence of static malicious adversaries if for every non-uniform PPT adversary $A$ for the real model, there exists a non-uniform PPT adversary $\simul$ for the ideal model, such that for every $I \subset [m]$\\
    \centerline{
    $
    \{\textsc{Ideal}_{\mathcal{F}, I, \simul(z)}(\bar x)
    \}_{\bar x, z}
    \overset{\mathrm{c}}{\equiv}
    \{\textsc{Real}_{\Pi, I, \adv(z)}(\bar x)
    \}_{\bar x, z}
    $
}
\\
where $\bar x = ( x_1, \ldots, x_m)$ and $x_i \in \{0,1\}^*$.
\label{def:mpc}
\end{definition}
Here, $\textsc{Ideal}_{\mathcal{F}, I, \simul(z)}(\bar x)$ denotes the joint output of the honest parties and $\simul$ from the ideal world execution of $\mathcal{F}$;
and
$\textsc{Real}_{\Pi, I, \adv(z)}(\bar x)$
denotes the joint output of the honest parties and $\adv$ from the real world execution of $\Pi$~\cite{Goldreich2004}.

As with malicious MPC, we cannot control what data a party chooses to input. The parties can, if they wish, augment the query to run tests on the input data (\eg interval checks). 
\sys also does not intend to maintain consistency of the datasets input by a party across \emph{different} queries as the dataset could have changed in the meantime.
If this is desired, \sys could in principle support this by writing multiple queries as part of a single bigger query, at the expense of performance.

Note that the query result might leak information about the underlying datasets,
and the parties should choose carefully what query results they are willing to share with each other. 
Alternatively, it may be possible to integrate techniques such as differential privacy~\cite{Dwork:DP, Flex:Johnson:VLDB} with \sys's MPC computation, to avoid leaking information about any underlying data sample; we discuss this aspect in more detail in \cref{s:discussion}.

\section{\sys's MPC decomposition protocol}\label{s:protocol}

In this section we present \sys's {\em secure MPC decomposition} protocol, the key enabler of our compiler's planning algorithm.

Our protocol may be of independent interest, and we present the cryptography in a self-contained way. 

Suppose that $m$ parties, $P_1,\ldots,P_m$, wish to securely compute a function $f$, represented by a circuit $C$, on their private inputs $x_i$. 
This can be done easily given a state-of-the-art MPC protocol
by having all the parties collectively evaluate the entire circuit using the protocol.
However, the key idea in \sys is that if $f$ can be ``nicely'' decomposed into multiple sub-circuits, we can achieve a protocol with a significantly better concrete efficiency, by having only a {\em subset} of the parties participate in the secure evaluation of each sub-circuit. %

\noindent For example, consider a function $f(x_1,\ldots,x_m)$ that can be evaluated by separately computing $y_1=h_1(x_1,\ldots,x_{i})$ on the inputs of parties $P_1 \dots P_i$, and $y_2=h_2(x_{i+1},\ldots,x_m)$ on the inputs of parties $P_{i+1} \dots P_m$,  followed by $\tilde f(y_1,y_2)$. 
That is,
$$
f(x_1,\ldots,x_m) = \tilde f\big( h_1(x_1,\ldots,x_{i}), h_2(x_{i+1},\ldots,x_m) \big).
$$

\noindent Such a decomposition of $f$ allows parties $P_1,\ldots,P_{i}$ to securely evaluate $h_1$ on their inputs (using an MPC protocol) and obtain output $y_1$.
In parallel, parties $P_{i+1},\ldots,P_m$ securely evaluate $h_2$ to get $y_2$. 
Finally, all parties securely evaluate $\tilde f$ on $y_1,y_2$ and obtain the final output $y$.
We observe that such a decomposition may lead to a more efficient protocol for computing $f$, since the overall communication and computation complexity of state-of-the-art concretely efficient MPC protocols (\eg \cite{WRK17:Global,Overdrive}) is at least quadratic in the number of involved parties. 
Furthermore, sub-circuits involving disjoint sets of parties can be evaluated in parallel.

Although appealing, this idea has some caveats:
\begin{compactenumerate}
    \item In a usual (``monolithic'') secure evaluation of $f$, the intermediate values $y_1,y_2$ remain secret, whereas the decomposition above reveals them to the parties as a result of an intermediate MPC protocol.
    \item Suppose that $h_1$ is a \emph{non-easily-invertible} function (\eg pre-image resistant hash function). 
        If all of $P_1,\ldots,P_{i}$ collude, they can pick an arbitrary ``output'' $y_1$, even without evaluating $h_1$, and input it to $\tilde f$.
        Since $h_1$ is  non-invertible, it is infeasible to find a pre-image of $y_1$; thus, such behavior is not equivalent to 
        the adversary's ability to provide an input of its choice
        (as allowed in the malicious setting). 
        In addition, such functions introduce problems in the proof's simulation as a PPT simulator cannot extract the corrupted parties' inputs with high probability.
        This attack, however, would not have been possible if $f$ had been computed entirely by all of $P_1,\ldots,P_m$ in a \emph{monolithic} MPC.
	\item If one party is involved in multiple sub-circuits and is required to provide the same input to all of them, then we have to make sure that its inputs are consistent.

\end{compactenumerate}
\noindent
In this section we show how to deal with the above problems, by
building upon the  MPC protocol of Wang \etal~\cite{WRK17:Global}.

First, we show how to securely transfer the output of one garbled circuit as input to a subsequent garbled circuit, an action called {\em soldering} (\cref{s:protocol:soldering}).
Our soldering is inspired by previous soldering techniques proposed in the MPC literature~\cite{NielsenO09,FrederiksenJNNO13,FrederiksenJNT15,KolesnikovNRTT17,AfsharHMR15,GargLOS15,HazayY16,Garg0MP16,LuO17,KellerY18,arx,arxrange}.
Here, we make the following contributions.
To the best of our knowledge, \sys is the first work to design a soldering technique for the state-of-the-art %
protocol of Wang \etal~\cite{WRK17:Global}. 
More importantly, whereas previous uses of soldering were limited to cases in which the {\em same set of parties} participate in both circuits, we show how to solder circuits when the first set of parties {\em is a subset of} the set of parties involved in the second circuit. 
This property is crucial for the performance of the individual sub-circuits in our overall protocol, as most of them can now be evaluated by non-overlapping subsets of parties, in parallel.

Second, as observed above, the decomposition of a function for MPC cannot be arbitrary. We therefore formalize the class of decompositions that are \emph{admissible} for MPC (\cref{s:protocol:mpctree}). Informally, we require that every sub-computation evaluated by less than $m$ parties must be efficiently invertible.
This fits the ability of a malicious party to choose its input before providing it to the computation.

Furthermore, we define the admissible circuit structures to be \emph{trees} rather than directed acyclic graphs.
That is, the function's decomposition may only take the form of a tree of sub-computations, and not an arbitrary graph. This is because if a node provides input to more than one parent node and all the parties at the node are corrupted, they may collude to provide inconsistent inputs to the different parents.
We therefore circumvent this input consistency problem by restricting valid decompositions to trees alone.
Even so, as we show in later sections, this model fits SQL queries particularly well, since many SQL queries can be naturally expressed as a tree of operations.

\subsection{Background}
\label{s:protocol:wrk}
\label{s:background}

We start by briefly introducing the cryptographic tools that our MPC protocol builds upon.
In particular, we build upon the maliciously-secure garbled circuit protocol of Wang \etal~\cite{WRK17:Global} (hereafter referred to as the WRK protocol).

\paragraph{Information-theoretic MACs (IT-MACs)}
IT-MACs~\cite{NNOB12:TinyOT} enable a party $P_j$ to authenticate a bit held by another party $P_i$.
Suppose $P_i$ holds a bit $x \in \{0,1\}$, and $P_j$ holds a key
$\Delta_j \in \{0,1\}^\kappa$ (where $\kappa$ is the security parameter).
$\Delta_j$ is called a \emph{global key} and $P_j$ can use it to authenticate multiple bits across parties.
Now, for $P_j$ to be able to authenticate $x$, $P_j$ is given a random {\em local key} $K_j[x] \in \{0,1\}^\kappa$ and $P_i$ is given the corresponding MAC tag $M_j[x]$ such that:
$$
M_j[x] = K_j[x]\oplus x \Delta_j.
$$
$P_j$ does not know the bit $x$ or the MAC, and $P_i$ does not know the keys; thus, $P_i$ can later reveal $x$ and its MAC to $P_j$ to prove it did not tamper with $x$.
In this manner, $P_i$'s bit $x$ can be authenticated to more than one party---each party $j$ holds a global key $\Delta_j$ and local key for $x$, $K_j[x]$. $P_i$ holds all the corresponding MAC tags $\{M_j[x]\}_{j\neq i}$.
We write $[x]^i$ to denote such a bit where $x$ is known to $P_i$, and is authenticated to {\em all} other parties.
Concretely, $[x]^i$ means that $P_i$ holds $(x, \{M_j[x]\}_{j\ne i})$, and every other party $P_j\neq P_i$ holds $K_j[x]$ and $\Delta_j$. 

Note that $[x]^i$ is XOR-homomorphic: given two authenticated bits $[x]^i$ and $[y]^i$, it is possible to compute the authenticated bit $[z]^i$ where $z = x \oplus y$ by simply having each party compute the XOR of the MAC / keys locally.

\paragraph{Authenticated secret shares}
In the above construction, $x$ is known to a single party and authenticated to the rest.
Now suppose that $x$ is {\em shared} amongst all parties such that no subset of parties knows $x$. In this case, each $P_i$ holds $x^i$ such that $x=\oplus_{i} x^i$.
To authenticate $x$, we can use IT-MACs on each share $x^i$ and distribute the authenticated shares $[x^i]^i$.
We write $\authshare{x}_\Delta$ to denote the collection of authenticated shares $\{ [x^i]^i \}_{i}$ 
under the global keys $\Delta=\{\Delta_i\}_{i}$. 
We omit the subscript in $\authshare{x}_\Delta$ if the global keys are clear from context. One can show that $\authshare{x}$ is XOR-homomorphic, \ie given $\authshare{x}$ and $\authshare{y}$ the parties can locally compute $\authshare{z}$ where $z=x\oplus y$.

\paragraph{Garbled circuits and the WRK protocol}
Garbled circuits~\cite{Yao86:GC,BHR12,BMR90} are a commonly used cryptographic primitive in MPC constructions.
Formally, an $m$-party garbling scheme is a pair of 
algorithms $(\fgarble,\feval)$
that allows a secure evaluation of a (typically Boolean) circuit $C$.
To do so, the parties first invoke $\fgarble$ with $C$, and obtain a garbled circuit $\garbled(C)$ and some extra information (each party may obtain its own secret extra information). Then, given the input $x_i$ to party $P_i$, the parties invoke $\feval$ with $\{x_i\}_i$ and obtain the evaluation output $y$.
(This is a simplification of a garbling scheme in many ways, but this abstraction suffices to understand the WRK protocol below.)
Typically, constructions utilizing a garbling scheme are in the {\em offline-online} model, in which they may invoke $\fgarble$ offline when they agree on the circuit $C$, and only later  they learn their inputs $\{x_i\}_i$ to the computation.

The WRK protocol \cite{WRK17:Global} %
is the state-of-the-art garbled circuit protocol that is maliciously-secure even when $m-1$ out of $m$ parties are corrupted. WRK follows the same abstraction described above, with its own format for a garbled circuit; thus, we denote its garbling scheme by $(\fwrkgarble,\allowbreak \fwrkeval)$. 
Our construction does not modify the inner workings of the protocol; therefore, we describe only its input and output layers, but elide internal details for simplicity.
\begin{mydescription}
	\item[$\fwrkgarble$:] 
		Given a Boolean circuit $C$, the protocol outputs a garbled circuit $\garbled(C)$. 
    The garbling scheme authenticates the circuit by maintaining IT-MACs on all input/output wires,\footnote{In fact, it does so for all the wires in the circuit; we omit this detail as we focus on the input / output interface.} as follows.
		Each party $P_i$ obtains a global key $\Delta_i$ for the circuit. In addition, each wire $w$ in the circuit is associated with a random ``masking'' bit $\lambda_w$ which is output to the parties as $\authshare{\lambda_w}_{\Delta}$.

	\item[$\fwrkeval$:] 
		The protocol is given a garbled circuit $\garbled(C)$. Then, for a party $P_i$ who wishes to input $b_w$ to input wire $w$, we have the parties input $\hat b_w=b_w\oplus \lambda_w$ instead; in addition, instead of receiving the real output bit $b_v$ the parties receive a masked bit $\hat b_v=b_v\oplus \lambda_v$.  %
		Note that $\lambda_w$ and $\lambda_v$ should be kept secret from the parties (except from the party who inputs $b_w$ or receives $b_v$, respectively). 
    The procedures by which parties privately translate masked values to real values and vice versa are simple and not part of the core functionality, as we describe below. 
\end{mydescription}

Using the above abstractions, the overall WRK protocol is simple and can be described as follows:
\begin{compactenumerate}
	\item {\it Offline.}
		The parties invoke $\fwrkgarble$ on $C$ and obtain $\garbled(C)$ and $\authshare{\lambda_w}$ for every input/output wire $w$.
	\item {\it Online.}
		\begin{compactenumerate}
			\item {\it Input.}
				If an input wire $w$ is associated with party $P_i$, who has the input bit $b_w$,  then the parties reconstruct $\lambda_w$ to $P_i$. Then, $P_i$ broadcasts the bit $\hat b_w=b_w\oplus\lambda_w$.
			\item {\it Evaluation.}
				The parties invoke $\fwrkeval$ on $\garbled(C)$ and the bit $\hat b_w$ for every input wire $w$. They obtain a bit $\hat b_v=b_v\oplus\lambda_v$ for every output wire $v$.
			\item {\it Output.}
				To reveal bit $b_v$ of an output wire $v$, the parties publicly reconstruct $\lambda_v$ and compute $b_v=\hat b_v\oplus \lambda_v$.
		\end{compactenumerate}
\end{compactenumerate}

\subsection{Soldering wires of WRK garbled circuits}\label{s:protocol:soldering}
The primary technique in \sys is to securely transfer the {\em actual value} that passes through an output wire of one circuit, without revealing that value, to the input wire of another circuit. This action is called \emph{soldering} \cite{NielsenO09}. We observe that the WRK protocol enjoys the right properties that enable soldering of its wires {\em almost for free}. In addition, we show how to extend the soldering notion even to cases where the set of parties who are engaged in the `next' circuit is a superset of the set of parties engaged in the current one.
This was not known until now. 
We believe this extension is of independent interest and may have  more applications beyond \sys.

Specifically, we wish to securely transfer the (hidden) output $b_v = \hat b_v \oplus \lambda_v$ on output wire $v$ of $\garbled(C_1)$ to the input wire $u$ of $\garbled(C_2)$. `Securely' means that $b_v=b_u$ should hold while keeping both $b_u$ and $b_v$ secret from the parties.
To achieve this, the parties need to securely compute the masked value of the input to the next circuit, as expected by the WRK protocol:
\begin{align*}
    \hat b_u = \lambda_u \oplus b_u 
             = \lambda_u \oplus b_v %
             = \lambda_u \oplus \lambda_v \oplus \hat b_v
\end{align*} 
and input it to $\fwrkeval$ for the next circuit.

Note that the parties already hold the three terms on the right hand side of the above equation---$\fwrkeval$ outputs $\hat b_v$ to the parties as a masked output when evaluating $\garbled(C_1)$, and the parties hold $\authshare{\lambda_v}$ and $\authshare{\lambda_u}$ as output from $\fwrkgarble$ on $C_1$ and $C_2$ respectively.
Thus,
one attempt to obtain $\hat b_u$ might be to have the parties compute the shares of $\authshare{\lambda_u \oplus \lambda_v \oplus \hat b_v}$ using XOR-homomorphism, and then publicly reconstruct it.
However, this operation is {\em not defined} unless the global key that each party uses in the constituent terms is the \emph{same}.
Since we 
do not modify the construction of $\fwrkgarble$ and $\fwrkeval$,
the global keys in the two circuits (and hence in $\authshare{\lambda_v}$ and $\authshare{\lambda_u}$) are different with high probability.

We overcome this limitation using the functionality $\fsolder$:
\begin{tcolorbox}[left=5pt,right=5pt,top=5pt,bottom=5pt]
    \small
    \begin{myfunctionality}
        \label{func:solder}
         $\fsolder(v,u)$ -- Soldering
    \end{myfunctionality}

    {\bf Inputs.}
    Parties in set $\partyset_1$ agree on $\hat b_v$ and have $\authshare{\lambda_v}_\Delta$ authenticated under global keys $\{\Delta_i\}_{i \in \partyset_1}$.
    Parties in set $\partyset_2$ (where $\partyset_1\subseteq \partyset_2$) have $\authshare{\lambda_u}_{\tilde \Delta}$ authenticated under global keys $\{\tilde \Delta_i\}_{i \in \partyset_2}$.     %

    \noindent
    {\bf Outputs.}
    Compute $\hat b_u = \lambda_u \oplus \lambda_v\oplus \hat b_v$. Then,
    \begin{compactitemize}
    	\item Output $\delta_i=\Delta_i\oplus\tilde\Delta_i$ for all $P_i\in\partyset_1$ to parties in $\partyset_1$.
        \item Output $\lambda_v^i \oplus \lambda_u^i$ for all $P_i\in\partyset_1$ to parties in $\partyset_1$.
    	\item Output $\lambda_u^i$ for all $P_i\in\partyset_2\setminus\partyset_1$ to everyone.
    	\item If $\authshare{\lambda_v}_\Delta$ and $\authshare{\lambda_u}_{\tilde\Delta}$ are valid then output $\hat b_u$ to parties in $\partyset_2$.
    	\item Otherwise, output $\hat b_u$ to the adversary and $\bot$ to the honest parties.
    \end{compactitemize}

\end{tcolorbox}

\noindent
Before proceeding, note that $\fsolder$ satisfies our needs: $\partyset_1$ and $\partyset_2$ are engaged in evaluating garbled circuits $\garbled(C_1)$ and $\garbled(C_2)$ respectively. $v$ is an output wire of $\garbled(C_1)$, and $u$ is an input wire of $\garbled(C_2)$. The parties in $\partyset_2$ want to transfer the actual value that passes through $v$, namely $b_v$, to $\garbled(C_2)$. That is, they want the actual value that would pass through $u$ to be $b_v$ as well. However, they do not know $b_v$, but only the masked value $\hat b_v$. Thus, by using $\fsolder$, they can obtain exactly what they need in order to begin evaluating $\garbled(C_2)$ with $b_u=b_v$.

    Along with the soldered result $\hat b_u$, functionality $\fsolder$ also reveals additional information to the parties---specifically, the values of $\delta_i$ (for all $P_i\in\partyset_1$); $\lambda_v^i \oplus \lambda_u^i$ (for all $P_i\in\partyset_1$); and $\lambda_u^i$ (for all $P_i\in\partyset_2\setminus\partyset_1$).
We model this extra leakage in the functionality as this information is revealed by our protocol that instantiates $\fsolder$.
However, we will show that this does not affect the security of our overall MPC protocol.

\paragraph{Instantiating $\fsolder$}
We start by defining a procedure for XOR-ing authenticated shares under \emph{different} global keys, which we denote $\boxplus$. 
That is, %
$\authshare{x}_{\Delta}\boxplus\authshare{y}_{\tilde\Delta}$ outputs $\authshare{x\oplus y}_{\tilde\Delta}$.

We observe that it is possible to implement $\boxplus$ in a very simple manner:
every party $P_i$ only needs to broadcast the difference of the two global keys: $\delta_i=\Delta_i\oplus\tilde \Delta_i$. 
Using this, the parties can switch the underlying global keys of $\authshare{x}$ from $\Delta_i$ to $\tilde\Delta_i$ by having each party $P_i$ compute new authentications of $x^i$, denoted $M^\prime_j[x^i]$, as follows. 
For every $j\neq i$, $P_i$ computes 
\begin{align*}
    M^\prime_j[x^i] &= M_j[x^i] \oplus x^i \delta_j \\
                    &= K_j[x^i] \oplus x^i \Delta_j \oplus x^i \delta_j 
                    = K_j[x^i] \oplus x^i \tilde \Delta_j
\end{align*}

So now, $x$ is shared and authenticated under the new global keys $\{\tilde \Delta_i \}_i$.
Given this procedure, we can realize $\fsolder$ as follows: the parties first compute $\authshare{b_v}_\Delta=\authshare{\lambda_v}_\Delta \oplus \hat b_v$; \footnote{XOR homomorphism works also when one literal is a constant, rather than an authenticated sharing.}
the parties then compute $\authshare{\hat b_u}_{\tilde\Delta} = \authshare{b_v}_\Delta\boxplus\authshare{\lambda_u}_{\tilde \Delta}$, and reconstruct $\hat b_u$ by combining their shares. 

Note that the description above (implicitly) assumes that $\partyset_1=\partyset_2$; however, if $\partyset_1\subset \partyset_2$ then the $\boxplus$ protocol does not make sense because parties in $\partyset_2$ that are not in $\partyset_1$ do not have a global key $\Delta_i$ corresponding to $\authshare{x}_\Delta$.
Forcing them to participate in the $\boxplus$ protocol with $\Delta_i=0$ would result in a complete breach of security as it would reveal $\delta_i=\Delta_i \oplus \tilde\Delta_i = \tilde\Delta_i$, which must remain secret!
We resolve this problem in the protocol $\psolder$ (\cref{prot:solder}) which extends $\boxplus$ to the case where $\partyset_1\subset \partyset_2$.

\begin{figure*}
\begin{tcolorbox}[left=5pt,right=5pt,top=5pt,bottom=5pt]
    \small
    \begin{myprotocol}
        \label{prot:solder}
        $\psolder$ -- Soldering
    \end{myprotocol}
    Denote by $\authshare{\lambda_u^{\partyset_1}}_{\tilde\Delta}$ the authenticated secret shares of $\lambda_u$ held by parties in $\partyset_1$ only. That is $\lambda_u^{\partyset_1} = \bigoplus_{i:P_i\in \partyset_1} \lambda_u^i$.
    \begin{compactenumerate}%
        \item 
            The parties in ${\partyset_1}$ reconstruct $\authshare{\hat b_u^{\partyset_1}}_{\tilde\Delta} = ( \hat b_v \oplus \authshare{\lambda_v}_{\Delta}) \boxplus \authshare{\lambda_u^{\partyset_1}}_{\tilde \Delta}$.
            
            Specifically, each party $P_i\in\partyset_1$ broadcasts:
            \begin{enumerate*}
            	\item the bit $\hat b_u^i=\lambda_v^i \oplus \lambda_u^i$, and
            	\item the difference $\delta_i=\Delta_i \oplus \tilde \Delta_i$ .
            \end{enumerate*}
        After receiving $\hat b_u^j$ and $\delta_j$ from every $P_j\in\partyset_1$, it computes
        \begin{eqnarray*}
        \hat b_u^{\partyset_1} &=& \hat b_v \oplus \bigoplus\nolimits_{i:P_i\in \partyset_1} \hat b_u^i,\\
        M_j[\hat b_u^i]&=&M_j[\lambda_v^i \oplus \lambda_u^i] = M_j[\lambda_v^i] \oplus M_j[\lambda_u^i] \oplus \lambda_v^i\cdot\delta_j
        = (K_j[\lambda_v^i]\oplus \lambda_v^i\cdot \Delta_j) \oplus (K_j[\lambda_u^i]\oplus \lambda_u^i\cdot \tilde \Delta_j) \oplus (\lambda_v^i\cdot\delta_j) \\
        &=& K_j[\lambda_v^i]\oplus K_j[\lambda_u^i]\oplus \lambda_v^i\cdot(\Delta_j\oplus \delta_j) \oplus \lambda_u^i\cdot \tilde \Delta_j
        = K_j[\lambda_v^i]\oplus K_j[\lambda_u^i]\oplus (\lambda_v^i\oplus \lambda_u^i)\cdot \tilde \Delta_j ~~~ \text{and} \\
        K_i[\hat b_u^j] &=& K_i[\lambda_v^j]\oplus K_i[\lambda_u^j]
        \end{eqnarray*}
 
        for every $j\in\partyset_1$ and broadcasts $M_j[\hat b_u^i]$.

        \item Parties $P_i \in \partyset_2 \setminus \partyset_1$ broadcast $\lambda_u^i$ and $M_j[\lambda_u^i]$ for all $j\in\partyset_2$.

        \item 
        	Parties $P_i\in\partyset_1$ verify that $K_i[\hat b_u^j] \oplus \hat b_u^j\cdot \tilde\Delta_i = M_i[\hat b_u^j]$ for all $j\in\partyset_1$.
        \item 
        	Parties $P_i\in\partyset_2$ verify that $K_i[\lambda_u^j] \oplus \lambda_u^j\cdot \tilde\Delta_i = M_i[\lambda_u^j]$ for all $j\in\partyset_2\setminus\partyset_1$.
        
        \item
            If verification fails, output $\bot$ and abort. Otherwise, output
            $$
            \hat b_u 
            =  \left(\bigoplus_{P_i\in\partyset_2} \lambda_u^i \right) \oplus b_u
            = \left(\bigoplus_{P_i\in\partyset_1} \lambda_u^i \right) \oplus \left(\bigoplus_{P_i\in\partyset_2\setminus\partyset_1} \lambda_u^i \right)  \oplus b_u
            = \hat b_u^{\partyset_1 } \oplus \left(\bigoplus_{P_i\in\partyset_2\setminus\partyset_1} \lambda_u^i \right) 
            $$
    \end{compactenumerate}

\end{tcolorbox}
\end{figure*}

\begin{theorem}
	\label{thm:soldering}
    Protocol $\psolder$ securely computes functionality $\fsolder$ (per \cref{def:mpc})
    in the presence of a static adversary that corrupts an arbitrary number of parties. %
\end{theorem}

\noindent We defer the proof to an extended version of our paper.

\subsection{Secure computation of circuit trees}
\label{s:protocol:mpctree}
Given a SQL query, \sys decomposes the query into a tree of circuits, where each \emph{non-root} node (circuit) in the tree involves only a subset of the parties.
We now describe how the soldering technique can be used to evaluate trees of circuits, while preserving the security of the overall computation. 
To this end, we first formalize the class of circuit trees that represent valid decompositions with respect to our protocol; then, we concretely describe our protocol for executing such trees.

We start with some preliminary definitions and notation. 
A {\em circuit tree} $T$ is a tree whose internal nodes are circuits, and the leaves are the tree's input wires (which are also input wires to some circuit in the tree).
Each node that provides input to an internal node $C$ in the tree is a child of $C$.
Since $T$ is a tree, this implies that all of a child's output wires may only be fed as input to a single parent node in the tree.

We denote a circuit $C$'s and a tree $T$'s input wires by $\inp(C)$ and $\inp(T)$ respectively.
Each wire $w\in \inp(T)$ is associated with one party $P_i$, in which case we write $\parties(w)=P_i$.
Let $G_1,\ldots,G_k$ be $C$'s children, we define 
$\parties(C)=\cup_{i=1}^k \parties(G_i)$. 
Note that we assume, without loss of generality, that the root circuit $C\in T$ has $\parties(C)=\{P_1,\ldots, P_m\}$ (\ie it involves inputs from all parties).
Our goal is to achieve secure computation for circuit trees; however, as discussed earlier, our construction does not support arbitrary trees. 
We now describe formally what can be achieved.

\begin{definition} \label{def:invertiblecircuit}
    A circuit $C: {\cal D} \to {\cal R}$ 
    (where ${\cal D}\subseteq \{0,1\}^k$ is $C$'s domain and ${\cal R}\subseteq \{0,1\}^\ell$ is the range)
    is {\em \sf invertible} if there is a polynomial time algorithm ${\cal A}$ (in the size of the circuit $|C|$) such that 
    given $y \in \{0,1\}^\ell$:
    $$
    {\cal A}(y) = 
    \begin{cases}
        x \text{ such that } x \in {\cal D} \text{ and } C(x) = y & \quad \text{if } y \in {\cal R}\\
        \perp & \quad \text{if } y \not\in {\cal R}
    \end{cases}
    $$

\end{definition}

Note that in the definition above, the circuit $C$ need not be ``full range'', \ie its range may be a subset of $\{0,1\}^\ell$. 
In such cases, we require that 
it is ``easy'' to verify that a given value $y\in\{0,1\}^\ell$ is also in ${\cal R}$. By easy we mean that it can be verified by a polynomial-size circuit. We also denote by ${\sf ver}_C(y)$ the circuit that checks whether a value $y\in\{0,1\}^\ell$ is in ${\cal R}$ and returns 0 or 1 accordingly.
Note that given a tree of circuits, the range of an intermediate circuit depends not only on the circuit's computation, but also on the ranges of its children because they limit the circuit's domain.
Thus, these ranges need to be deduced topologically for the tree, using which the ${\sf ver}_C$ circuit is manually crafted.

\begin{definition}\label{def:addmissibletrees}
	For $t<m$, the class of $t$-{\em \sf admissible circuit trees}, denoted $\mathcal{T}(t)$, contains all circuit trees $T$, such that $C$ is invertible for all $C\in T$ where $|\parties(C)|\leq t$. In addition, each circuit $C$ that is parent to circuits $G_1,\ldots,G_k$ has ${\sf ver}_{G_1},\ldots,{\sf ver}_{G_k}$ embedded within it as sub-circuits, 
    and $\parties(C)=\cup_{i=1}^k \parties(G_i)$. 
\end{definition}

The above suggests that there {\em may} indeed be non-invertible circuits (\eg a preimage resistant hash) in the tree; the only restriction is that such a circuit should be evaluated by more than $t$ parties.
The definition of MPC for circuit trees follows the general definition of MPC \cite{Goldreich2004}, as presented below. %

\begin{tcolorbox}[left=5pt,right=5pt,top=5pt,bottom=5pt]
	\small
	\begin{myfunctionality}
		\label{func:mpctree}
		$\fmpctree$ -- MPC for circuit trees
	\end{myfunctionality}
	{\bf Parameters.} A circuit tree $T$ and parties $P_1,\ldots,P_m$.\\ %
    {\bf Inputs.} For each $w\in\inp(T)$ where $P_i=\parties(w)$, wait for an input bit $b_w$ from $P_i$.\\
    {\bf Outputs.} The bit $b_w$ for every $w$ in $T$'s output wires, given by evaluating $T$ in a topological order from leaves to root.
\end{tcolorbox}

We realize $\fmpctree$ using the protocol $\pmpctree$ (\cref{prot:mpctree}), which is our overall protocol for securely executing circuit trees. 
\begin{tcolorbox}[float,left=5pt,right=5pt,top=5pt,bottom=5pt]
    \small
    \begin{myprotocol}
        \label{prot:mpctree}
        $\pmpctree$ - MPC for circuit trees
    \end{myprotocol}
    {\bf Parameters.}
    The circuit tree $T$. Parties $P_1,\ldots,P_m$.

    {\bf Inputs.}
    For $w\in\inp(T)$, $P_i=\parties(w)$ has $b_w\in\{0,1\}$.

    {\bf Protocol.}
    \begin{compactenumerate}%
    \item {\it Offline.}
        For every circuit $C\in T$, $\parties(C)$ run $\fwrkgarble(C)$ to obtain $\garbled(C)$
        along with  $\authshare{\lambda_w}$ for all input and output wires $w$.
    \item {\it Online.} 
        For each circuit $C$ in $T$ (topologically) do:
        \begin{compactenumerate}
        \item {\it Input.}
            For every $u\in\inp(C)$:
            If $u\in\inp(T)$ and $P_i=\parties(u)$ then $\parties(C)$ reconstruct $\lambda_u$ to $P_i$.
            Else, if $u$ is connected to an output wire $v$ of a child circuit $C^\prime$ then run $\fsolder(v,u)$, by which $\parties(C)$ obtain $\hat b_u$.
        \item {\it Evaluate.}
            Run $\fwrkeval$ on $\garbled(C)$ and $\hat b_u$ for every $u\in\inp(C)$, by which $\parties(C)$ obtain $\hat b_v$ for every $C$'s output wire $v$. 
            If $G_1,\ldots,G_c$ are $C$'s children then abort if an intermediate value ${\sf ver}(G_i)=0$ for some $i\in[c]$.
        \item {\it Output.}
            If $C$ is the root of $T$, reconstruct $\authshare{\lambda_w}$ for every $w\in\outp(C)$, by which all parties obtain $b_w = \hat w\oplus \lambda_w$.
        \end{compactenumerate}
    \end{compactenumerate}
\end{tcolorbox}
The protocol works as follows. 
In the offline phase the parties simply garble all circuits using $\fwrkgarble$; each circuit is garbled independently from the others. 
Then, beginning from the tree's leaf nodes, the parties evaluate the circuits using $\fwrkeval$, such that each circuit $C$ is evaluated only by $\parties(C)$ (not all the parties).
When a value on an output wire of some circuit $C^\prime$ should travel privately to the input wire of the next circuit $C$ then $\parties(C)$ run the soldering protocol. 
As discussed above, $\parties(C^\prime)$ may be a subset of $\parties(C)$. 
Once all the nodes have been evaluated, the parties operate exactly as in the WRK protocol in order to reveal the actual value on the output wire. %

We prove the security of protocol $\pmpctree$ per the following theorem in 
an extended version of our paper.
We remark that our protocol inherits the random oracle assumption from its use of the WRK protocol.

\begin{theorem}\label{th:security:mpc}
	Let $t<m$ be the number of parties corrupted by a static adversary.
    Then, protocol $\pmpctree$ securely computes $\fmpctree$ (per \cref{def:mpc}) for any $T\in\mathcal{T}(t)$, in the random oracle model and the $\fsolder$-hybrid model. %
\end{theorem}

We stress that intermediate values (output wires of internal nodes) are authenticated secret shares, each using fresh randomness, and thus kept secret from the adversary. In particular, the adversary's input is independent of these values.

Note that by our construction, if there is a sub-tree rooted at a circuit $C$ such that $\parties(C)$ are all corrupted, then the adversary may skip the `secure computation' of that sub-tree and simply provide inputs directly to $C$'s parent. This, however, does not form a security issue because
a malicious adversary may change its input anyway, and
the sub-tree is invertible---hence,
whatever input is given to $C$'s parent, it can be used to extract {\em some} possible adversary's input to the tree's input wires (and hence to the functionality) that leads to the target output from the functionality.

In the following sections, we describe how \sys executes SQL queries by transforming them into circuit trees that can be securely executed using our protocol.

\section{\sys's circuit primitives}\label{s:primitives}

\sys executes a query by first representing it as a tree of Boolean circuits, and then processing the circuit tree using its efficient MPC protocol. 
To construct the circuits, \sys uses a small set of circuit primitives which we describe in turn.
In later sections, we describe how \sys composes these primitives to represent SQL operations and queries.

\subsection{Filtering}\label{s:primitives:filter}
Our first building block is a simple circuit ($\filter$) that takes a list of elements as input, and passes each element through a sub-circuit that compares it with a specified constant.
If the check passes, it outputs the element, else it outputs a zero. 

\subsection{Multi-way set intersection}\label{s:primitives:psi}

Next, we describe a circuit for computing a \emph{multi-way} set intersection.
Prior work has mainly focused on designing Boolean circuits for two-way set intersections~\cite{PSI:HEK12:NDSS,SetOps:Blanton:2012}; here we design optimized circuits for intersecting multiple sets.
Our circuit extends the two-way SCS circuit of Huang \etal~\cite{PSI:HEK12:NDSS}. %
We start by providing a brief overview of the SCS circuit, and then describe how we extend it to multiple sets.

\parheadskip{The two-way set intersection circuit ($\twopsi$)}\label{s:primitives:psi:two}
The sort-compare-shuffle circuit of Huang \etal~\cite{PSI:HEK12:NDSS} takes as input two sorted lists of size $n$ each with \emph{unique} elements, and outputs a list of size $n$ containing the intersection of the lists interleaved with zeros (for elements that are not in the intersection).
(1)~The circuit first merges the sorted lists. (2)~Next, it filters intersecting elements by comparing adjacent elements in the list, producing a list of size $n$ that contains all filtered elements interleaved with zeros.
(3)~Finally, it shuffles the filtered elements to hide positional information about the matches.

In \sys's use cases, set intersection results are often not the final output of an MPC computation, and are instead intermediate results upon which further computation is performed.
In such cases, the shuffle operation is not performed.

\parheadskip{A multi-way set intersection circuit ($\mpsi$)}\label{s:primitives:psi:multi}
Suppose we wish to compute the intersection over three sets $A, B$ and $C$.
A straightforward approach is to compose two $\twopsi$ circuits together into a larger circuit (\eg as $\twopsi(\twopsi(A, B), C)$).
However, such an approach doesn't work out-of-the-box because the intermediate output $O=\twopsi(A, B)$ needs to be sorted before it can be intersected with $C$, as expected by the next $\twopsi$ circuit. 
While one can accomplish this by sorting the output, 
it comes at the cost of an extra $O(n \log^2 n$) gates.

Instead of performing a full-fledged sort, 
we exploit the observation that, essentially, the output $O$ of $\twopsi$ is the \emph{sorted} result of $A \cap B$ interleaved with zeros.
So, we transform $O$ into a \emph{sorted multiset} via an intermediate \emph{monotonizer} circuit $\mono$ that replaces each zero in $O$ with the nearest preceding non-zero value.
Concretely, given $O = (a_1 \ldots a_n)$ as input, $\mono$ outputs $M = (b_1 \ldots b_n)$, such that $b_i = a_i$ if $a_i \ne 0$, else $b_i = b_{i-1}$.
For example, if $O = (1, 0, 2, 3, 0, 4)$, then $\mono$ converts it to $M = (1, 1, 2, 3, 3, 4)$.

Since $M$
now also contains duplicates, 
for correctness of the overall computation, 
the next $\twopsi$ that intersects $M$ with $C$ needs to be able to discard these duplicates. 
We therefore modify the next $\twopsi$ circuit: (i)~the circuit tags a bit to each element in the input lists that identifies which list the element belongs to,
\ie it appends 0 to every element in the first list, and 1 to every element in the second; 
(ii)~the comparison phase of the circuit additionally verifies that elements with equal values have different tags.
These modifications ensure that duplicates in the same intermediate list aren't added to the output.
We refer to this modified $\twopsi$ circuit as $\twopsitagged$.

The described approach generalizes to multiple input sets in an identical manner.
Note that in general, there can be many ways of constructing the binary tree of $\twopsi$ circuits (\eg a left-deep vs.\ balanced tree).
In \cref{s:planning} we describe how \sys's compiler picks the optimal design when executing queries.

\subsection{Multi-way sort}\label{s:primitives:sort}

Given $m$ sorted input lists of size $n$ each, a multi-way sort circuit $\msort$ merges the lists into a single sorted list of size $m \times n$, 
using a binary tree of bitonic merge operations (implemented as the $\merge$ circuit).

\subsection{Multi-way set union}\label{s:primitives:psu}

Our next building block is a circuit for multi-way set unions.
In designing the circuit, we extend the two-way set union circuit of Blanton and Aguiar~\cite{SetOps:Blanton:2012}.

\parheadskip{The two-way set union circuit ($\twopsu$)}\label{s:primitives:psu:two}
Given two sorted input lists of size $n$ each with unique elements, the $\twopsu$ circuit produces a list of size $2n$ containing the set union of the inputs.
Blanton and Aguiar~\cite{SetOps:Blanton:2012} proposed a
$\twopsu$ circuit similar to $\twopsi$:
\begin{enumerate*}[label=(\arabic*)]
    \item It first merges the input lists into a single sorted list.
    \item Next, it removes duplicate elements from the list: for every two consecutive elements $e_{i}$ and $e_{i+1}$, if $e_{i} \ne e_{i+1}$ it outputs $e_i$, else it outputs 0. %
    \item Finally, the circuit randomly shuffles the filtered elements to hide positional information.
\end{enumerate*}

\parheadskip{A multi-way set union circuit ($\mpsu$)}\label{s:primitives:psu:multi}
It might be tempting to construct a multi-way set union circuit by composing multiple $\twopsu$ circuits together, similar to $\mpsi$.
However, such an approach is sub-optimal: unlike the intersection case where intermediate lists remain size $n$, in unions the intermediate result size grows as more input lists are added.
This leads to an unnecessary duplication of work in subsequent circuits. 
Instead, we construct a multi-way analogue of the $\twopsu$ circuit, as follows:
(1)~We first merge all $m$ input lists together into a single sorted list using an $\msort$ circuit. (2)~We then remove duplicate elements from the sorted list, in a manner identical to $\twopsu$.
We refer to the de-duplication sub-circuit in $\mpsu$ as $\dedup$.
The $\mpsu$ circuit may thus alternately be expressed as a composition of circuits: $\dedup\circ\msort$.

\subsection{Input verification}\label{s:primitives:verify}
Our description of the circuits thus far ($\mpsi$, $\mpsu$, and $\msort$) assumes that their inputs  are sorted.
While this assumption is safe in the case of semi-honest adversaries, it fails in the presence of malicious adversaries who may arbitrarily deviate from the MPC protocol.
For malicious security, we need to additionally \emph{verify} within the circuits that the inputs to the circuit are indeed sorted sets.
To this end, we augment the circuits with \emph{input verifiers} $\ver$, that scan each input set comparing adjacent elements $e_i$ and $e_{i+1}$ in pairs to check if $e_{i+1} > e_i$ for all $i$; if so, it outputs a 1, else 0.
When a given circuit is augmented with input verifiers, it additionally outputs a logical \code{AND} over the outputs of all constituent $\ver$ circuits.
This enables all parties involved in the computation to verify that the other parties did not cheat during the MPC protocol.

\section{Decomposable circuits for SQL operators}\label{s:sql}

Given a SQL query, \sys decomposes it into a tree of SQL operations and maps individual operations to Boolean circuits.
For some operations---namely, joins, group-by, and order-by operations---the Boolean circuits can be {\em further decomposed} into a tree of sub-circuits, which results in greater efficiency.
In this section, %
we show how \sys expresses individual SQL operations as circuits using the primitives described in \cref{s:primitives}, decomposing the circuits further when possible.
Later in \cref{s:planning}, we describe the overall algorithm 
for transforming queries into circuit trees and executing them using our MPC protocol.

\parheadskip{Notation}
We express \sys's transformation rules using traditional relational algebra~\cite{Codd:relational}, augmented with the notion of parties to capture the collaborative setting.
Let $\{P_1,\ldots,P_m\}$ be the set of parties in the collaboration.
Recall that we write $\Reln{R}{P_i}$ to denote a relation $R$ (\ie a set of rows) held by $P_i$.
We also repurpose $\Union$ to denote a simple concatenation of the inputs, as opposed to the set union operation.
The notation for the remaining relational operators are as follows: $\Select$ filters the input; $\Sort$ performs a sort; $\Join$ is an equijoin; and $\Groupby$ is group-by.

\subsection{Joins}\label{s:sql:join}

Consider a collaboration of $m$ parties, where each party $P_i$ holds a relation $R_i$ and wishes to compute an $m$-way join: %
$$
\Join(\Reln{R_1}{P_1}, \ldots, \Reln{R_m}{P_m})
$$
\sys converts \emph{equijoin} operations---joins conditioned on an equality relation between two columns---to set intersection circuits.
Specifically, \sys maps an $m$-way equijoin operation
to an $\mpsi$ circuit. 
For all other types of join operations, such as joins based on column comparisons or compound logical expressions, \sys expresses the join using a simple Boolean circuit that performs a series of operations per pairwise combination of the inputs.
However, a recent study~\cite{Flex:Johnson:VLDB} notes that the vast majority of joins in real-world queries (76\%) are equijoins. Thus, a majority of join queries can benefit from our optimized design of set intersection circuits.

\parheadskip{Decomposing joins across parties}
If parties don't care about privacy, the simplest way to execute the join would be to perform a series of 2-way joins in the form of a tree.
 For example, one way to evaluate a $4$-way join is to order the constituent joins as $((R_1 \Join R_2) \Join (R_3 \Join R_4))$.
To mimic this decomposition, \sys starts by designing an $\mpsi$ Boolean circuit to compute the operation (with $m=4$).
\sys then evaluates the $\mpsi$ circuit by decomposing it into its constituent sub-circuits as follows:
\begin{compactenumerate}
    \item First, each party locally sorts its input sets (as required by the $\mpsi$ circuit).
    \item Next, parties $P_1$ and $P_2$ jointly compute a $\twopsi$ operation over $R_1$ and $R_2$, followed by the monotonizer $\mono$.
    In parallel, parties $P_3$ and $P_4$ compute a similar circuit over $R_3$ and $R_4$.
    The $\twopsi$ circuits are augmented with $\ver$ sub-circuits that verify that the input sets are sorted.
    
    \item Finally, all four parties evaluate a $\twopsitagged$ circuit over the outputs of the previous step; %
        as before, the circuit includes a $\ver$ sub-circuit to check that the inputs are sorted.
    Note that though the evaluated circuit takes two sets as input, the circuit computation involves all four parties.
\end{compactenumerate}

\noindent
In general, multiple tree structures are possible for decomposing an $m$-way join.
\sys's compiler (which we describe in \cref{s:planning}) derives the best plan for the query using a cost model.

\parheadskip{Joins over multisets}
\sys's $\mpsi$ circuit can be extended to support joins over multisets in a straightforward manner.
We defer the details to an extended version of our paper.

\subsection{Order-by limit}
\label{s:sql:orderby}

In the collaborative setting, the $m$ parties may wish to perform an order-by operation (by some column $c$) on the union of their results, optionally including a limit $l$:
$$
\Sort_{c, l}(\Union_i \Reln{R_i}{P_i})
$$
\sys maps order-by operations directly to the $\msort$ circuit. 
If the operation includes a limit $l$, then the circuit only outputs the wires corresponding to the first $l$ results.

Recall from \cref{s:primitives:sort} that $\msort$ is a composition of $\merge$ sub-circuits (that perform bitonic merge operations).
If the operation includes a limit $l$, then we make an optimization that reduces the size of the overall circuit.
We note that 
since the circuit's output only contains wires corresponding to the first $l$ elements of the sorted result, any gates that do not impact the first $l$ elements can be discarded from the circuit.
Hence, if an element is outside the top $l$ choices for any intermediate $\merge$, then we discard the corresponding gates.

\parheadskip{Decomposing order-by across parties}
Since the $\msort$ circuit is composed of a tree of $\merge$ sub-circuits, it can be straightforwardly decomposed across parties by distributing the constituent $\merge$ sub-circuits.
For example, one way to construct a $4$-party sort circuit is: $\merge(\merge(R_1, R_2), \allowbreak \merge(R_3, R_4))$.
To decompose this:
\begin{compactenumerate}
\item Each party first sorts their input locally (as expected by the $\msort$ circuit).
\item Parties $P_1$ and $P_2$ compute a $\merge$ sub-circuit; $P_3$ and $P_4$ do the same in parallel.
\item All $4$ parties finally $\merge$ the previous outputs. 
\end{compactenumerate}
Once again, multiple tree structures are possible for distributing the $\merge$ circuits, and the \sys compiler's planning algorithm picks the best structure based on a cost model.

\subsection{Group-by with aggregates}\label{s:sql:groupby}
Suppose the parties wish to compute a group-by operation over the union of their relations (on some column $c$), followed by an aggregate $\Sigma$ per group:
$$
\Groupby_{c,\Agg}(\Union_i \Reln{R_i}{P_i})
$$

\noindent
\sys starts by mapping the operator to a $\Sigma\circ\mpsu$ circuit that computes the aggregate function $\Agg = \code{SUM}$.
To do so, we extend the $\mpsu$ circuit with support for aggregates.
Recall from \cref{s:primitives:psu} that the $\mpsu$ circuit is a composition of sub-circuits $\dedup\circ\msort$.

Let the input to the group-by operation be a list of tuples of the form $t_i = (a_i, b_i)$, such that the $a_i$ values represent the columns over which groups are made, and the $b_i$ values are then aggregated per group.
\begin{compactenumerate}
    \item In the $\msort$ phase, \sys evaluates the $\msort$ sub-circuit over the $a_i$ values per tuple, while ignoring $b_i$.
    \item In the $\dedup$ phase, for every two consecutive tuples $(a_i, b_i)$ and $(a_{i+1}, b_{i+1})$, the circuit outputs $(a_i, b_i)$ if $a_{i} \ne a_{i+1}$, else it outputs $(0, b_i)$
    \item In addition, we augment the $\dedup$ phase to compute aggregates over the $b_i$ values.
The circuit makes another pass over the tuples $(a^\prime_i, b_i)$ output by $\dedup$ while maintaining a running aggregate \code{agg}: if $a^\prime_i = 0$ then it updates \code{agg} with $b_i$ and outputs $(0, 0)$; otherwise, it outputs $(a^\prime_i, \code{agg})$.
\end{compactenumerate}

\parheadskip{Decomposing group-by across parties}
\sys decomposes group-by operations in two ways.
First, group-by operations with aggregates can typically be split into two parts: local aggregates per party, followed by a joint group-by aggregate over the union of the results.
This is a standard technique in database theory.
For example, suppose $\Agg = \code{COUNT}$. In this case, the parties can first compute local counts per group, and then evaluate a joint sum per group over the local results.
Rewriting the operation in this manner helps \sys reduce the amount of joint computation performed using a circuit, and is thus beneficial for performance.

Second, 
we note that the joint group-by computation can be further decomposed across parties.
Specifically, 
the $\msort$ phase of the overall $\mpsu$ circuit (as described above) can also be distributed across the parties in a manner identical to order-by (as described in \cref{s:sql:orderby}).

\subsection{Filters and Projections}\label{s:sql:select:project}
Filtering is a common operation in queries (\ie the \code{WHERE} clause in SQL), and parties in a collaboration may wish to compute a filter on the union of their input relations:
$$
\Select_f(\Union_i \Reln{R_i}{P_i})
$$
where $f$ is the condition for filtering.
\sys maps the operation to a $\filter$ circuit.
Filtering operations at the start of a query can be straightforwardly distributed by evaluating the filter locally at each party, before performing the union.

As regards projections, typically, these operations simply exclude some columns from the relation. 
Given a relation, \sys performs a projection by simply discarding the wires corresponding to the non-projected columns.

\section{Query execution}\label{s:planning}

\begin{figure*}
    \centering
    \includegraphics[width=\textwidth]{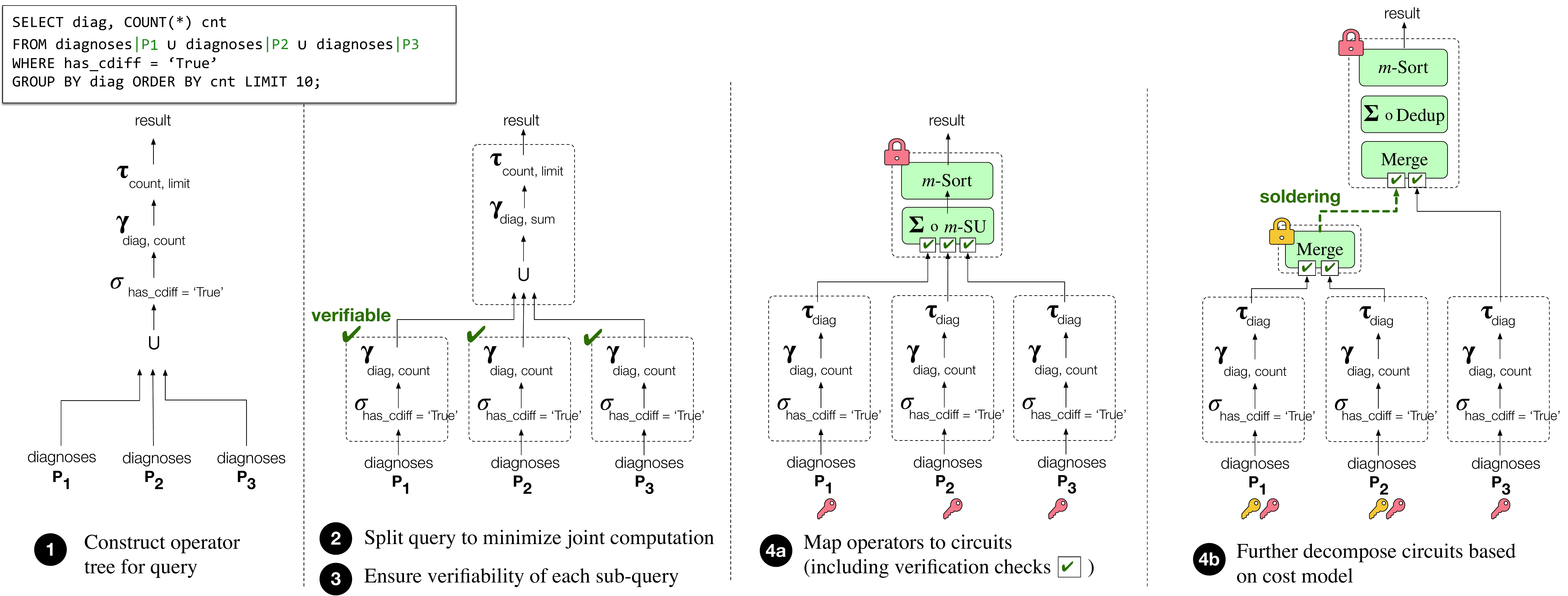}
    \caption{Query execution in \sys. Colored keys and locks indicate which parties are involved in which MPC circuits.}
    \label{fig:query}
\end{figure*}

We now describe how \sys executes a query by decomposing it into a tree of circuits.
In doing so, \sys's compiler ensures that the resulting tree satisfies the requirements of our MPC protocol (per \cref{def:addmissibletrees})---namely, that each circuit in the tree is invertible.

\subsection{Query decomposition and planning}
We start by describing the \sys compiler's query decomposition algorithm.
\label{s:planning:algorithm}
Given a query, the compiler transforms the query into a circuit tree in four steps, as illustrated in \cref{fig:query}.
We use the medical query from \cref{s:overview} as a running example.

\paragraph{Step \bubble{1}: Construction of tree of operators}
\sys first represents the query as a \emph{tree} of relational operations.
The leaves of the tree are the input relations of individual parties, and the root outputs the final query result.
Each non-leaf node represents an operation that will be jointly evaluated only by the parties whose data the node takes as input. %
Thus, the set of parties evaluating a node is always a superset of its children.

While a query can naturally be represented as a directed acyclic graph (DAG) of relational operators, \sys recasts the DAG into a tree to satisfy the \emph{input consistency} requirements of our MPC protocol.
Specifically, \sys ensures that the outputs of no intermediate node (or the input tables at the leaves) are fed to more than one parent node.
This is because in such cases, if any two parents are evaluated by disjoint sets of parties, then this leads to a potential input inconsistency---that is, if all the parties at the current node collude, then there is no guarantee that they provide the same input to both parents.
A tree representation resolves this problem.

\cref{fig:query} illustrates the query tree for the medical query and comprises the following sequence of operator nodes---the input tables of the parties (in the leaves) are first concatenated into a single relation which is then processed jointly using a filter, a group-by aggregate, and an order-by limit operator.

\paragraph{Step \bubble{2}: Query splitting}\label{s:planning:local}
Next, \sys logically rewrites the query tree, splitting it such that the parties perform as much computation as possible locally over their plaintext data,
(\ie filters and aggregates), 
thereby reducing the amount of computation that need to be performed jointly using MPC.
To do so, it applies traditional relational equivalence rules that (i)~push down selections past joins and unions, and (ii)~decomposes group-by aggregates into local aggregates followed by a joint aggregate. %

For example, as shown in \cref{fig:query}, \sys rewrites the medical query in both these ways.
Instead of performing the filtering jointly (after concatenating the parties' inputs), \sys pushes down the filter past the union and parties apply it locally. In addition, it further splits the group-by aggregate---parties first compute local counts per group, and the local counts are jointly summed up to get the overall counts.

Though such an approach has also been explored in prior work~\cite{SMCQL:Bater:VLDB, Conclave:Volgushev:Eurosys}, an important difference in \sys is that while prior approaches assume a semi-honest threat model, \sys targets security against malicious adversaries who may arbitrarily deviate from the specified protocol.
To protect against malicious behavior, \sys's split is different than the semi-honest split; \sys performs two actions:
(i)~additionally verifies that all local computations are valid; and
(ii)~ensures that the splitting does not introduce input consistency problems.
We describe how \sys tackles these  issues next.

\paragraph{Step \bubble{3}: Verifying intermediate operations}\label{s:planning:verify}
We need to take a couple of additional steps before we can execute the tree of operations securely using our MPC protocol.
As \cref{s:protocol:mpctree} points out, to be maliciously secure, the tree of circuits needs to be ``admissible'' (per \cref{def:addmissibletrees}), \ie each intermediate operation in the tree must be invertible,  and each intermediate node must also be able to verify that the output produced by its children is possible given the query.

Thus, in transforming a query to a circuit tree, 
\sys's compiler deduces the set of outputs each intermediate operation can produce, while ensuring the operation is invertible.
For example, a filter of the type ``\code{WHERE 5 < age < 10}'' requires that in all output records, each value in column \code{age} must be between $5$ and $10$.
Note that the values of intermediate outputs also vary based on the set of preceding operations. 
For more complex queries, the constraints imposed by individual operators accumulate as the query tree is executed.

\sys's compiler traverses the query tree upwards from the leaves to the root, and identifies the constraints at every level of the tree.
For simplicity, we limit ourselves to the following types of constraints induced by relational operators: %
(i)~each column in a relation can have range constraints of the type $n_1 \le a \le n_2$, where $n_1$ and $n_2$ are constants;
(ii)~the records are ordered by a single column;
or 
(iii)~the values in a column are distinct. 
If the cumulative constraints at an intermediate node in the tree are limited to the above, then \sys's compiler marks the node as \emph{verifiable}.
If a node produces outputs with different constraints, then the compiler marks it as \emph{unverifiable}---for such nodes, \sys \emph{merges} the node with its parent into a single node and proceeds as before. %

If a node / leaf feeds input to more than one parent (perhaps as a result of the query rewriting in the previous step), then the compiler once again merges the node and all its parents into a single node, in order to avoid input consistency problems.

At the end of the traversal, the root node is the only potentially unverifiable node in the tree, but this does not impact security. Since all parties compute the root node jointly, the correctness of its output is guaranteed.

As an example, in \cref{fig:query}, the local nodes at every party locally evaluate the filter $\Select_{\tt has\_cdiff=True}$, which constrains the column \code{has\_cdiff} to the value \code{`True'}, and satisfies condition (i) above.
The subsequent group-by aggregate operation $\Groupby_{\tt diag,count}$ does not impose any constraint on either \code{diag} or \code{count} (since parties are free to provide inputs of their choice, assuming there are no constraints on the input columns). %
The local nodes are thus marked verifiable.
All remaining operations are performed jointly by all parties at the root node, and thus do not need to be checked for verifiability.

In 
our extended paper,
we work out in detail how \sys's compiler deduces the range constraints imposed by various relational operations (\ie what needs to be verified).
Then, we show the invertibility of relational operations given these constraints.
This ensures that the resulting tree is admissible, and satisfies the requirements of \sys's MPC protocol.

\paragraph{Step \bubble{4}: Mapping operators to circuits}\label{s:planning:circuits}
The final step is to map each jointly evaluated node in the query tree to a circuit (per \cref{s:sql}): $\Select$ maps to the $\filter$ circuit, $\Join$ maps to $\mpsi$, group-by aggregate maps to $\Agg\circ\mpsu$, and order-by-limit maps to $\msort$.
In doing so, 
\sys's compiler uses a planning algorithm that \emph{further decomposes} each circuit into a tree of circuits %
based on a cost model (described shortly).

For example, for the medical query in \cref{fig:query}, \sys maps the group-by aggregate operation $\Groupby_{\tt diag,sum}$ to a $\Sigma\circ\mpsu$ circuit.
Note that $\mpsu$ requires its inputs to be sorted; therefore, the compiler augments the children nodes with sort operations $\Sort_{\tt diag}$.
It then further decomposes the $\msort$ phase of $\mpsu$ into a tree of $\merge$ sub-circuits, per \cref{s:sql:groupby}.

This tree of circuits is finally evaluated securely using our MPC protocol.
Note that at each node, only the parties that provide the node input are involved in the MPC computation.

\subsection{Cost model for circuit decomposition}\label{s:planning:costmodel}
The planning algorithm models the latency cost of evaluating a circuit tree in terms of the constituent cryptographic operations.
It then enumerates possible decomposition plans, assigns a cost to each plan, and picks the optimal plan for decomposing the circuit.

Recall from \cref{s:protocol} that the cost of executing a circuit via MPC can be divided into
an offline phase (for generating the circuits), and an online phase (for evaluating the circuits).

Given a circuit tree $T$, let the root circuit be $C$ with children $C_0$ and $C_1$.
Let $T_0$ and $T_1$ refer to the subtrees rooted at nodes $C_0$ and $C_1$ respectively.
Then, \sys's compiler models the total latency cost $\cal C$ of evaluating $T$ as:
\begin{align*}
    {\cal C}(T) &= \max({\cal C}(T_0),{\cal C}(T_1))
                 +~\max({\cal C}_\mathsf{solder}(T_0),{\cal C}_\mathsf{solder}(T_1))\\
                &+~{\cal C}_\mathsf{offline}(C) + {\cal C}_\mathsf{online}(C)
\end{align*}
Essentially, since subtrees can be computed in parallel, the cost model counts the maximum of these two costs, followed by the cost of soldering the subtrees with the root node. 
It adds this to the cost of the offline and online phases for $T$'s root circuit $C$, ${\cal C}_\mathsf{offline}$  and ${\cal C}_\mathsf{online}$ respectively.

We break down each cost component in terms of two unit costs by examining the MPC protocol: the unit computation cost $L_s$ of performing a single symmetric key operation, and the unit communication cost $L_{i,j}$ (pairwise) between parties $P_i$ and $P_j$.
\sys profiles these unit costs during system setup.
In addition, the costs also depend on the size of the circuit being computed $|C|$ (\ie the number of gates in the circuit), the size of each party's input set $|I|$, and the number of parties $m$ computing the circuit. 
For simplicity, the analysis below assumes that each party has identical input set size; however, the model can be extended in a straightforward manner to accommodate varying input set sizes as well.

The soldering cost ${\cal C}_\mathsf{solder}$ can be expressed as $(m-1)|I|\cdot\max_{i,j}(L_{i,j})$ (since it involves a single round of communication between all parties).
Next, we analyze the WRK protocol to obtain the following equations:
$$
{\cal C}_\mathsf{offline}(C) = (m-1)|C|\cdot \max(L_{i,j}) + 4|C| \cdot L_s  + |C|\cdot\max(L_{1,i})
$$
In more detail, in the offline phase, each party (in parallel with the others) communicates with the $m-1$ other parties to create a garbled version of each gate in the circuit; each gate requires 4 symmetric key operations (one per row in the truth table representing the gate); they then send their individual garbled gates (in parallel) to the evaluator.
Our analysis here is a simplification in that we ignore the cost of some function-independent preprocessing steps from the offline phase. 
This is because these steps are independent of the input query, and thus do not lie in the critical path of query execution.

Similarly, the cost of the online phase can be expressed as
\begin{align*}
    {\cal C}_\mathsf{online}(C) = (m- & 1) |I|\cdot \max(L_{i,j}) \\
                                      & + (m-1)|I|\cdot\max(L_{1,i}) + (m-1)|C| \cdot L_s
\end{align*}
In this phase, the garblers communicate with all other parties to compute and send their encrypted inputs to the evaluator; in addition, the evaluator communicates with each garbler to obtain encrypted versions of its own inputs. The evaluator then evaluates the gates per party.
The size of the circuit $|C|$ depends on the function that the circuit evaluates (per \cref{s:primitives}), the number of inputs, and the bit length of each input.

\section{Evaluation}\label{s:evaluation}
In this section, we demonstrate \sys's improvements over 
running queries as monolithic cryptographic computations. 
We use vanilla AGMPC (with monolithic circuit execution) as the baseline. %
The highlights are as follows.
On the set of representative queries from \cref{s:api}, we observe runtime improvements of up to $10\times$ of \sys's building blocks, with a reduction in resource consumption of up to $11\times$. These results translate into runtime improvements of up to $10\times$ for the joint computation in the benchmarked queries.
\sys's query splitting technique provides a further improvement of up to $10\times$, bringing the net improvement to over $100\times$.
Furthermore, on the TPC-H analytics benchmark~\cite{TPCH:Web}, \sys's improvements range from $3\times$ to $145\times$.

\begin{figure}[t]
    \centering
    \begin{minipage}[b]{0.48\linewidth}
        \centering
        \includegraphics[width=\linewidth]{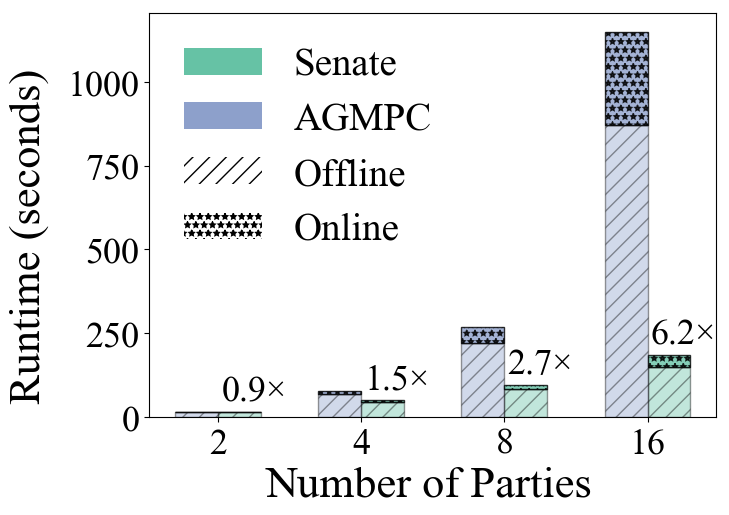}
        \subcaption{$\mpsi$ of $1K$ inputs/party.}
        \label{figure:psi_varying_parties}
    \end{minipage}
    \begin{minipage}[b]{0.48\linewidth}
        \centering
        \includegraphics[width=\linewidth]{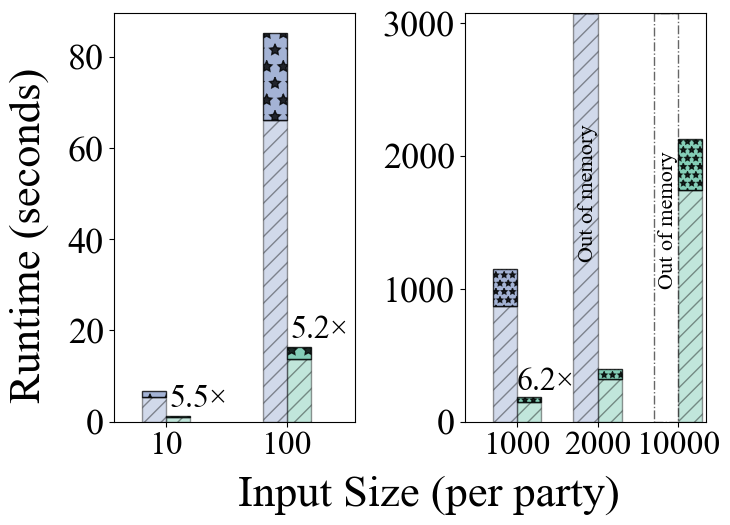}
        \subcaption{$\mpsi$ with $16$ parties.}
        \label{figure:psi_varying_input}
    \end{minipage}
    \caption{\small Performance of $\mpsi$ in LAN.}
    \label{figure:psi_performance}
\end{figure}

\begin{figure}[t]
    \centering
    \begin{minipage}[b]{0.48\linewidth}
        \centering
        \includegraphics[width=\linewidth]{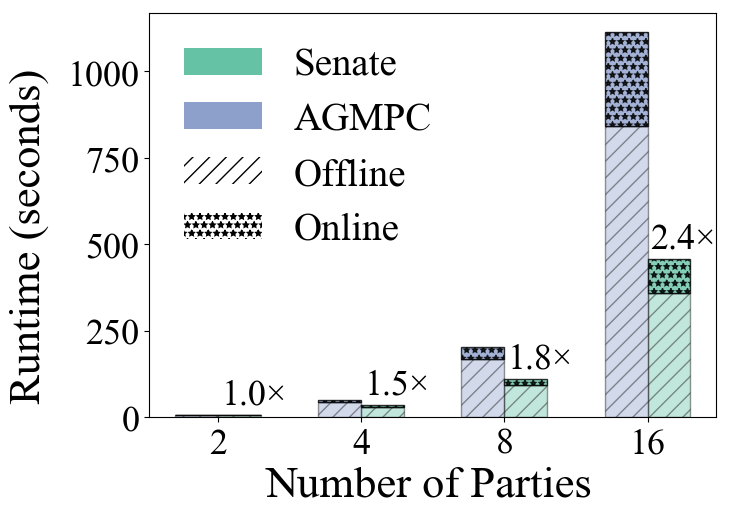}
        \subcaption{$\msort$ of $600$ inputs/party.}
        \label{figure:sort_varying_parties}
    \end{minipage}
    \begin{minipage}[b]{0.48\linewidth}
        \centering
        \includegraphics[width=\linewidth]{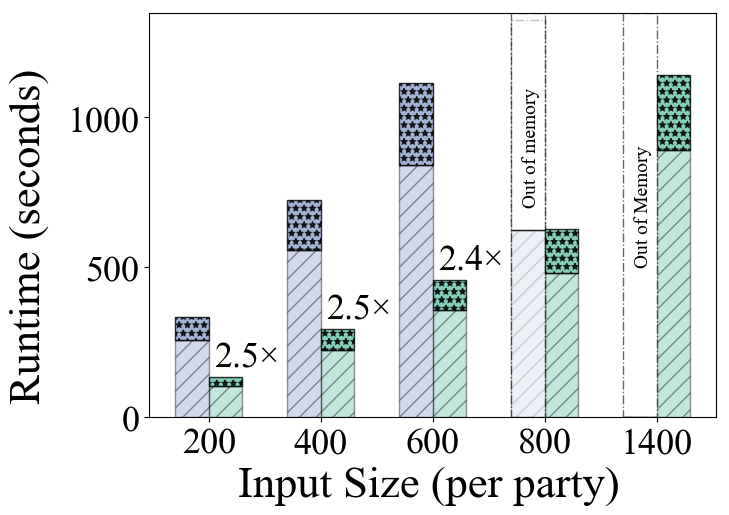}
        \subcaption{$\msort$ with $16$ parties.}
        \label{figure:sort_varying_input}
    \end{minipage}
    \captionsetup{skip=\dimexpr\abovecaptionskip-2pt}
    \caption{\small Performance of $\msort$ in LAN.}
    \label{figure:sort_performance}
\end{figure}

\begin{figure}[t]
    \centering
    \begin{minipage}[b]{0.48\linewidth}
        \centering
        \includegraphics[width=\linewidth]{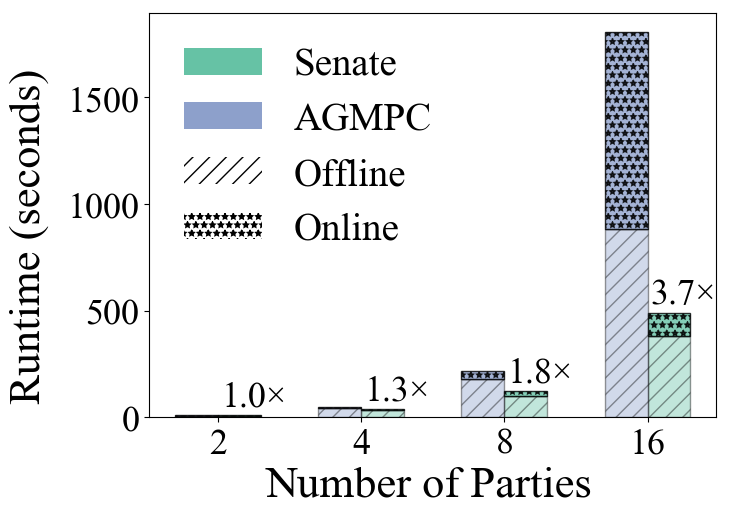}
        \subcaption{$\mpsu$ of $600$ inputs/party.}
        \label{figure:psu_varying_parties}
    \end{minipage}
    \begin{minipage}[b]{0.48\linewidth}
        \centering
        \includegraphics[width=\linewidth]{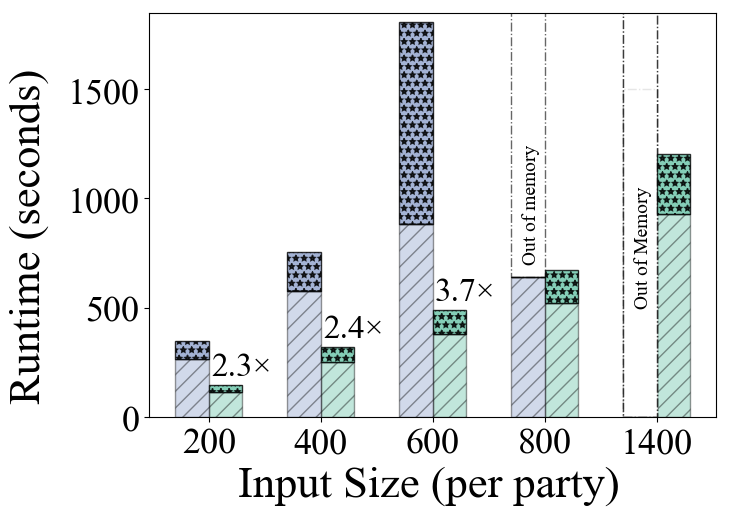}
        \subcaption{$\mpsu$ with $16$ parties.}
        \label{figure:psu_varying_input}
    \end{minipage}
    \caption{\small Performance of $\mpsu$ in LAN.}
    \label{figure:psu_performance}
\end{figure}

\begin{figure}[t]
    \centering
    \begin{minipage}[b]{0.48\linewidth}
        \centering
        \includegraphics[width=\linewidth]{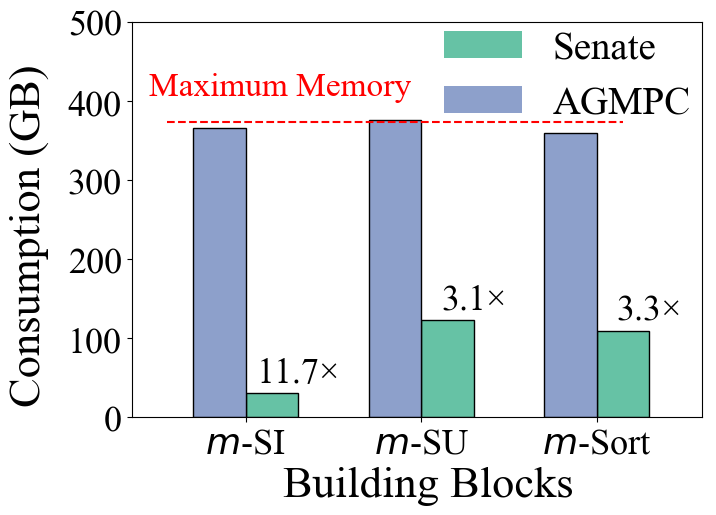}
        \subcaption{Peak memory usage}
        \label{figure:mem_consumption}
    \end{minipage}
    \begin{minipage}[b]{0.48\linewidth}
        \centering
        \includegraphics[width=\linewidth]{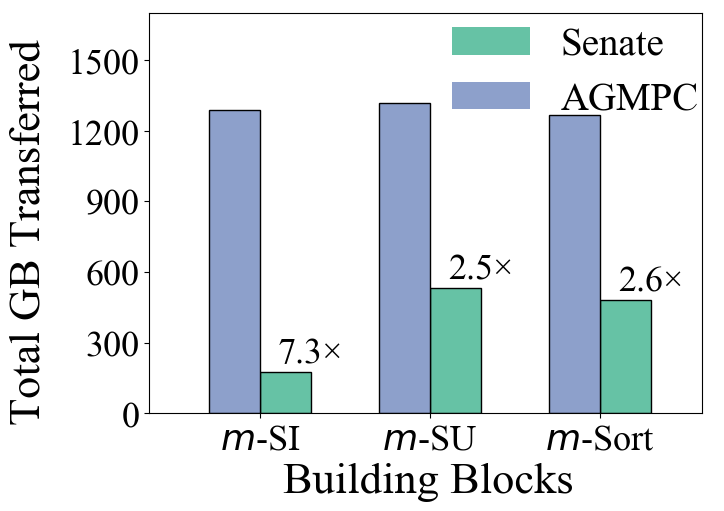}
        \subcaption{Network usage}
        \label{figure:net_consumption}
    \end{minipage}
    \captionsetup{skip=\dimexpr\abovecaptionskip-2pt}
    \caption{\small Resource consumption of building blocks (16 parties).}
    \label{figure:resource_consumption}
\end{figure}

\begin{figure*}[t!]
    \centering
    \begin{tabular}{cccc}
         \begin{minipage}[b]{0.23\linewidth}
             \centering
             \includegraphics[width=\linewidth]{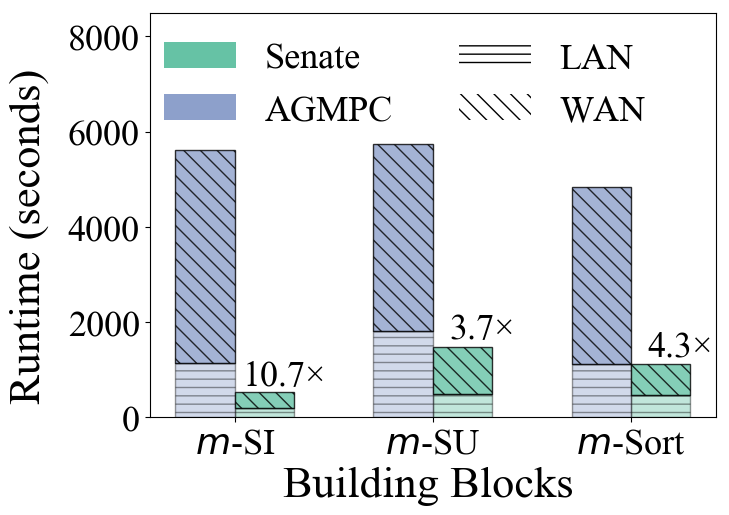}
             \caption{Building blocks in WAN.}
             \label{figure:performance_in_wan}
         \end{minipage}
         &
         \begin{minipage}[b]{0.23\linewidth}
             \centering
             \includegraphics[width=\linewidth]{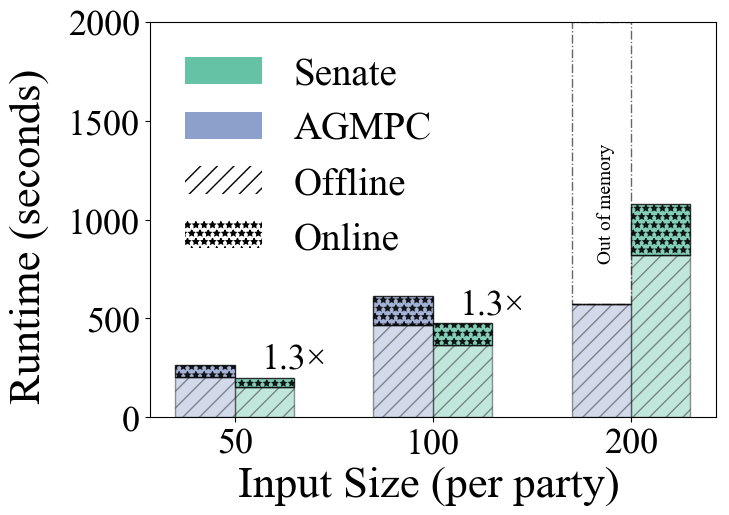}
             \caption{Query 1 with $16$ parties.}
             \label{figure:comorbidity_varying_input}
         \end{minipage}
         &
         \begin{minipage}[b]{0.23\linewidth}
             \centering
             \includegraphics[width=\linewidth]{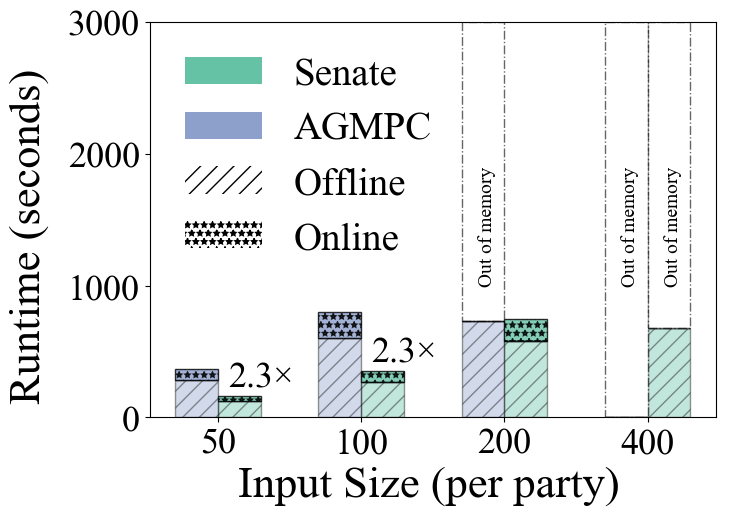}
             \caption{Query 2 with $16$ parties.}
             \label{figure:pass_varying_input}
         \end{minipage}
         &
         \begin{minipage}[b]{0.23\linewidth}
             \centering
             \includegraphics[width=\linewidth]{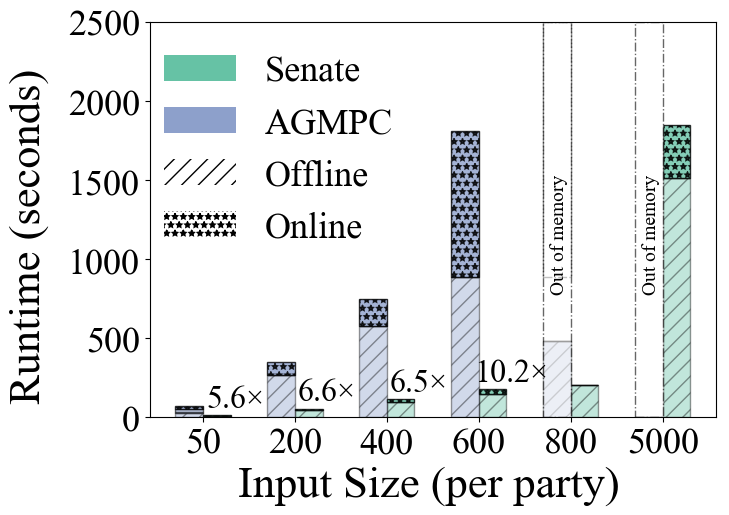}
             \caption{Query 3 with $16$ parties.}
             \label{figure:cc_varying_input}
         \end{minipage}
    \end{tabular}
    \label{figure:performance_2}
\end{figure*}

\begin{figure}[t]
    \centering
    \begin{tabular}{cc}
        \begin{minipage}[b]{0.46\linewidth}
            \centering
            \includegraphics[width=\linewidth]{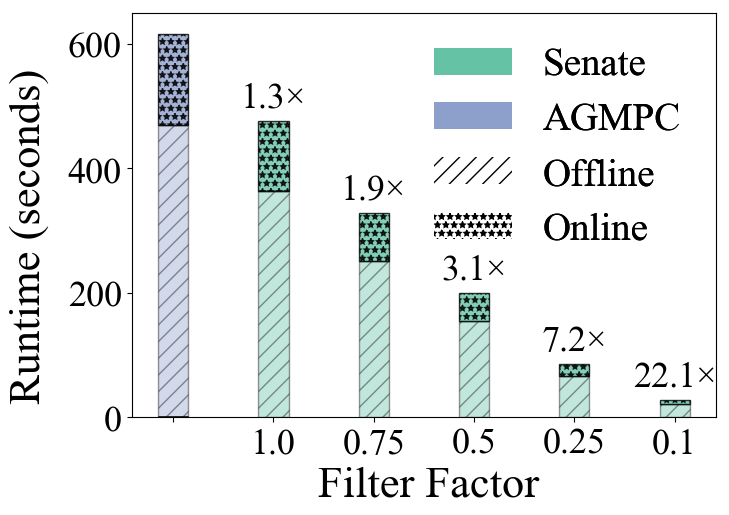}
            \subcaption{Query 1 with $100$ inputs/party.}
            \label{figure:comorbidity_varying_sensitivity}
        \end{minipage}
        &
        \begin{minipage}[b]{0.46\linewidth}
            \centering
            \includegraphics[width=\linewidth]{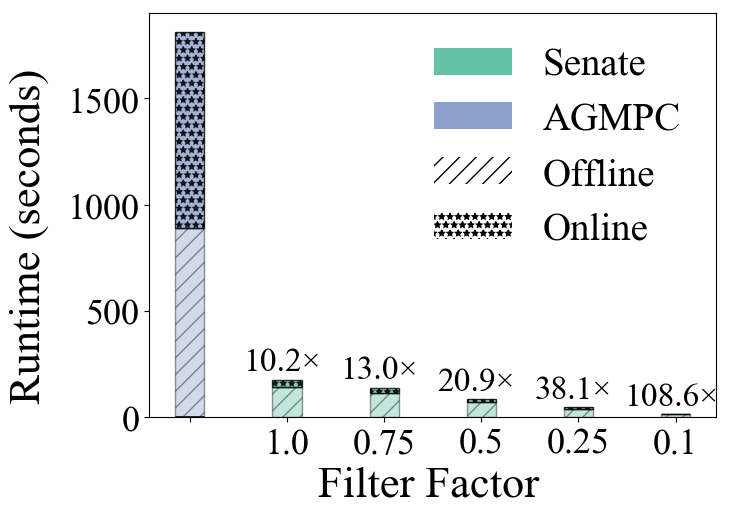}
            \subcaption{Query 3 with $600$ inputs/party.}
            \label{figure:cc_varying_sensitivity}
        \end{minipage}
    \end{tabular}
    \captionsetup{skip=\dimexpr\abovecaptionskip-5pt}
    \caption{\small Effect of query splitting on runtime. }
    \label{figure:performance_3}
\end{figure}

\begin{figure}[t]
    \centering
    \begin{minipage}[b]{0.48\linewidth}
        \centering
        \includegraphics[width=\linewidth]{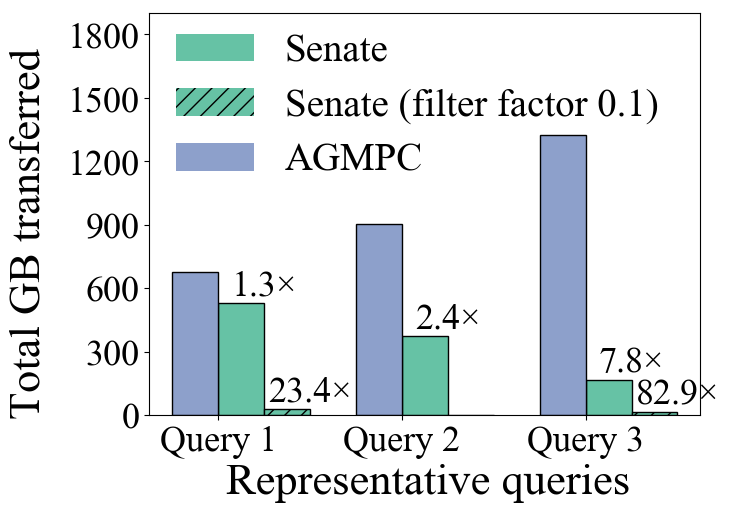}
        \caption{Network usage.}
        \label{figure:queries_comm}
    \end{minipage}
    \hfill
    \begin{minipage}[b]{0.48\linewidth}
        \centering
        \includegraphics[width=\linewidth]{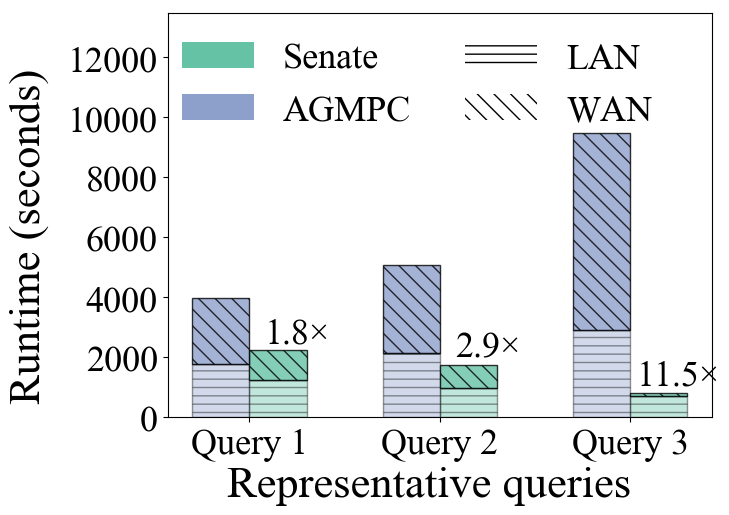}
        \caption{Queries in WAN.}
        \label{figure:queries_wan}
    \end{minipage}
\end{figure}

\parheadskip{Implementation}\label{s:implementation}
We implemented \sys on top of the AGMPC framework\cite{AGMPC:Git}, a state-of-the-art implementation of the WRK protocol~\cite{WRK17:Global} for $m$-party garbled circuits with malicious security.
Our compiler works with arbitrary bit lengths for inputs; in our evaluation, 
we set the data field size to be  integers of 32 bits, unless otherwise
specified.

\parheadskip{Experimental Setup}\label{s:eval:setup}
We perform our experiments using r5.12xlarge Amazon EC2
instances in the Northern California region. Each instance offers 
$48$ vCPUs and \SI{384}{\giga\byte} of RAM, and was additionally provisioned with \SI{20}{\giga\byte} of 
swap space, to account for transient spikes in memory requirements. 
We allocated similar instances in 
the Ohio, Northern Virginia and Oregon regions for wide-area network
experiments.

\hyphenation{AG-MPC}

\subsection{\sys's building blocks}\label{ss:building_block_eval}
We evaluate \sys's building blocks described in
\cref{s:primitives}---$\mpsi$, $\msort$, and $\mpsu$. For each building block, we compare 
the runtimes of each phase of the computation of 
\sys's efficient primitives to a similar implementation of the operator 
as a single circuit in both LAN and WAN settings (\cref{figure:psi_performance,figure:psu_performance,figure:sort_performance}, and \cref{figure:performance_in_wan}). We observe substantial improvements for our 
operators owing to reduced number of parties evaluating each sub-circuit and the
evaluation of various such circuits in parallel (per \cref{s:sql}).
We also measure the improvement in resource consumption due to \sys in \cref{figure:resource_consumption}. 

\parheadskip{Multi-way set intersection circuit ($\mpsi$)} We compare the evaluation time of  
an $\mpsi$ circuit across $16$ parties with varying input sizes in 
\cref{figure:psi_varying_input} and observe runtime improvements ranging from 
$5.2\times$--$6.2\times$. 
This is because our decomposition enables the input size
to stay constant for each sub-computation, allowing us to reduce the input set
size to the final $16$-party computation. Note that, while \sys can compute
a set intersection of $10K$ integers, AGMPC is unable to compute it for $2K$ integers, and
runs out of memory during the offline phase. 
\cref{figure:psi_varying_parties,figure:performance_in_wan}
plot the runtime of a circuit with varying number of parties in
LAN and WAN settings respectively, and observe an improvement of 
up to $10\times$. 
This can be similarly attributed to our decomposable circuits, 
which reduce the data transferred across all the parties, leading to 
significant improvements in the WAN setting.

\cref{figure:mem_consumption,figure:net_consumption} plot 
the trend of the peak memory and total network consumption of \sys
compared to AGMPC with $1K$ integers across varying number
of parties.

\parheadskip{Multi-way Sort circuit ($\msort$)} \cref{figure:sort_varying_parties,figure:sort_varying_input} illustrate the runtimes of a sorting
circuit with varying number of parties and varying input sizes respectively.
We observe that \sys's implementation is up to $4.3\times$ 
faster for $16$ parties, and can scale to twice as many inputs as AGMPC. This
is also corroborated by the $3.3\times$ reduction in peak memory requirement for $600$ integers
and \approx\SI{780}{\giga\byte} reduction in the amount of data transferred, as shown in 
\cref{figure:mem_consumption,figure:net_consumption}.

\parheadskip{Multi-way set union circuit ($\mpsu$)} \cref{figure:psu_varying_input} 
plots the runtime of a set union circuit with varying input sizes and $16$ parties.
As discussed in~\cref{s:primitives}, an $\mpsu$ circuit can be expressed as 
$\dedup\circ\msort$. 
Hence, we expect to trends similar to 
the $\msort$ circuit. However, we observed a stark increase in runtime
for the single circuit evaluation of $600$ integers across $16$ parties due to
the exhaustion of the available memory in the system and subsequent use of 
swap space (see \cref{figure:mem_consumption}). We observe a 
similar trend in \cref{figure:psu_varying_parties,figure:performance_in_wan}.

\subsection{End-to-end performance}\label{ss:end_to_end_eval}
\subsubsection{Representative queries}
We now evaluate the performance of \sys on the three representative queries
discussed in \cref{s:api} with a varying number of parties (\cref{figure:cc_varying_input,figure:comorbidity_varying_input,figure:pass_varying_input}).
In addition, we quantify the benefit of \sys's query splitting for different filter factors, \ie the fraction of inputs filtered as a result of any local computation (\cref{figure:performance_3}).
We also measure the total network usage of the queries in \cref{figure:queries_comm}; and \cref{figure:queries_wan} plots the performance of the queries in a WAN setting.

\parheadskip{Query 1 (Medical study)} \cref{figure:comorbidity_varying_input}
plots the runtime of \sys and AGMPC on the medical example query with varying
input sizes. Note that, the input to the circuit for a query consists of 
all the values in the row required to compute the final result. We observe 
a performance improvement of $1.3\times$ for an input size of $100$ rows, and are 
also able to scale to higher input sizes.
\cref{figure:comorbidity_varying_sensitivity} illustrates the benefit of 
\sys's consistent and verified query splitting for different filter 
factors. We compare the single circuit implementation of the query for
$100$ inputs per party, and are able to achieve a runtime improvement of $22\times$
for a filter factor of $0.1$.
The improvement in network consumption follows a similar trend, reducing usage by \approx$23\times$ with a filter factor of 0.1 (\cref{figure:queries_comm}).

\parheadskip{Query 2 (Prevent password reuse)} \cref{figure:pass_varying_input}
plots the runtime of \sys and AGMPC with varying input sizes. Each row in this
query consists of a $32$ bit user identifier, and a $256$ bit password hash.
Since the query involves a group-by with aggregates, which is mapped to an
extended $\mpsu$ (per~\cref{s:primitives}), we observe a trend %
similar to \cref{figure:psu_varying_input}. 
We remark that this query does not benefit from \sys's query splitting.

\parheadskip{Query 3 (Credit scoring)} We evaluate the third query with $16$
parties and varying input sizes in~\cref{figure:cc_varying_input}, and observe
that \sys is $10\times$ faster than AGMPC for $600$ input rows, and is able to scale
to almost $10$ times the input size. The introduction of a local filter into the query, 
with a filter factor of 0.1 reduces the runtime by $100\times$. We attribute this to our efficient $\mpsi$
implementation which optimally splits the set intersection and
parallelizes its execution across parties.
The reduction in network usage (\cref{figure:queries_comm}) is also similar.

\smallskip
\noindent
In the WAN setting, the improvement in query performance with \sys largely mimics the LAN setting; \cref{figure:queries_wan} plots the results in the absence of query splitting (\ie filter factor of 1).
Overall, we find that \sys MPC decomposition protocol alone improves performance by up to an order of magnitude over the baseline. In addition, \sys's query splitting technique can further improve performance by another order of magnitude, depending on the filter factor.

\subsubsection{TPC-H benchmark}
\label{s:eval:tpch}
To stress test \sys on more complex query structures, 
we repeat the performance experiment by evaluating \sys on the TPC-H benchmark~\cite{TPCH:Web}, an industry-standard analytics benchmark.
The benchmark comprises a rich set of 22 queries on data split across 8 tables.
The query structures are complex: for example, query 5 involves 5 joins across 6 tables, several filters, cross-column multiplications, aggregates over groups, and a sort.
Existing benchmarks for analytical queries (including TPC-H) have no notion of collaborations of parties, so we created a multi-party version of TPC-H by assuming that each table is held by a different party.

\begin{figure}[t]
    \centering
    \includegraphics[width=\linewidth]{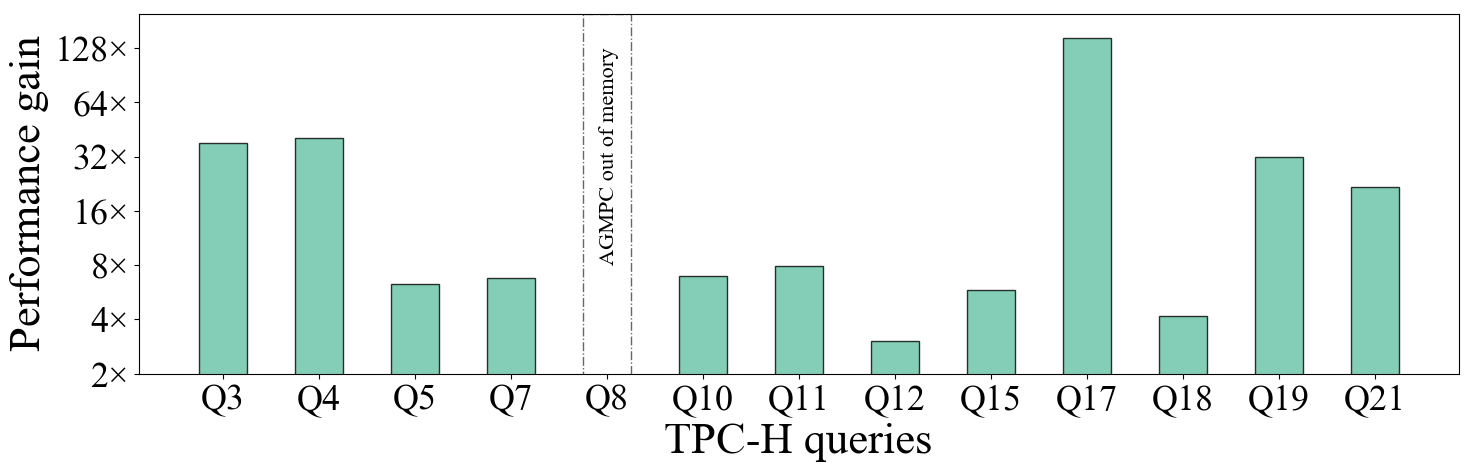}
    \caption{\sys's performance on TPC-H queries.}
    \label{fig:tpch}
\end{figure}

We measure \sys's performance on 13 out of these 22 queries; the other queries are either single-table queries,
or perform operations that \sys currently does not support (namely, substring matching, regular expressions, and UDFs). %
For parity, we assume 1$K$ inputs per party across all queries, and a filter factor of 0.1 for local computation that results from \sys's query splitting.
\cref{fig:tpch} plots the results.
Overall, \sys improves performance by $3\times$ to $145\times$ over the AGMPC baseline across 12 of the 13 queries; query 8 runs out of memory in the baseline.

\begin{figure}[t]
    \centering
    \begin{minipage}[b]{0.48\linewidth}
        \centering
        \includegraphics[width=\linewidth]{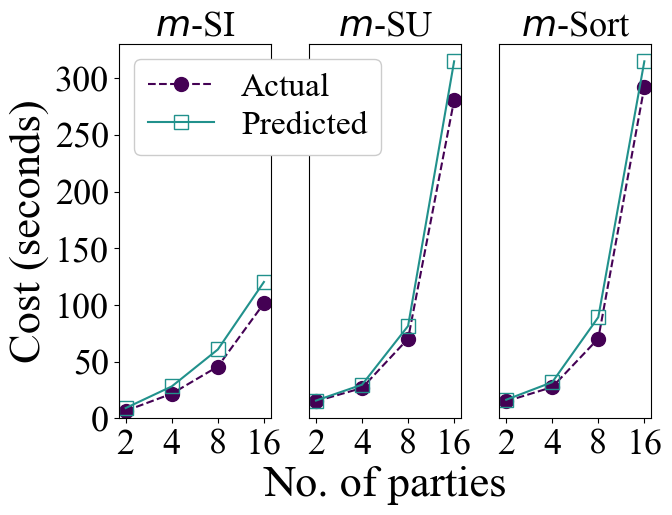}
        \caption{Accuracy of cost model.}
        \label{fig:cost_model}
    \end{minipage}
    \hfill
    \begin{minipage}[b]{0.48\linewidth}
        \centering
        \includegraphics[width=\linewidth]{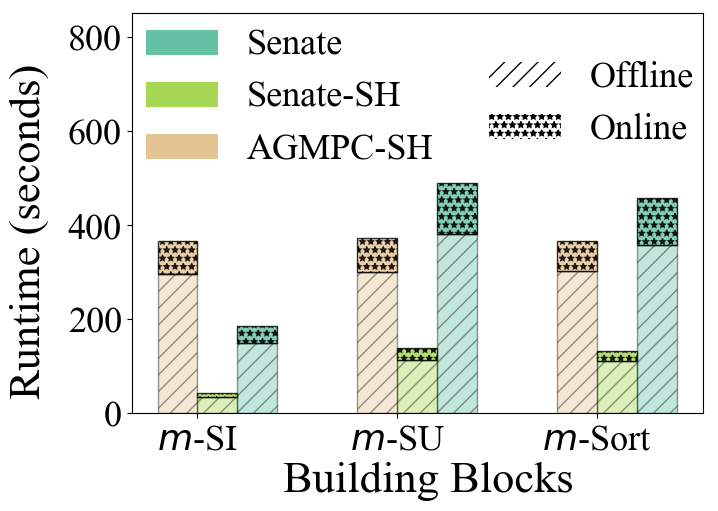}
        \caption{Semi-honest baselines}
        \label{figure:semi_honest_building_blocks}
        \label{figure:semi_honest_performance}
    \end{minipage}
\end{figure}

\subsection{Accuracy of \sys's cost model}
\label{s:eval:costmodel}
We evaluate our cost model (from \cref{s:planning:costmodel}) using \sys's circuit primitives.
We compute the costs predicted by the cost model for the primitives, and compare them with the measured cost of an actual execution. 
As detailed in \cref{s:planning:costmodel}, the cost model does not consider the function independent computation in the offline phase of the MPC protocol as it does not lie in the critical path of query evaluation; we therefore ignore the function independent components from the measured cost.
\cref{fig:cost_model} shows that our theoretical cost model approximates the actual costs well, with an average error of \approx20\%.

\subsection{\sys versus other protocols}
\label{ss:eval:relatedwork}
\parhead{Custom PSI protocols}
There is a rich literature on custom protocols for PSI operations.
While custom protocols are faster than general-purpose systems like \sys, their functionality naturally remains limited.
We quantify the tradeoff between generality and performance by comparing \sys's PSI cost to that of custom PSI protocols.
We compare \sys with the protocol of Zhang \etal~\cite{Zhang:CCSW}, a state-of-the-art protocol for multiparty PSI with malicious security.\footnote{We note that the protocol of Zhang \etal provides malicious security only against adversaries that do not simultaneously corrupt two parties, while \sys is secure against arbitrary corruptions. However, the only custom protocols we're aware of that tolerate arbitrary corruptions (for more than two parties) either rely on expensive public-key cryptography (and are slower than general-purpose MPC, which have improved tremendously since these proposals)~\cite{DMRY11,CJS12}, or do not provide an implementation~\cite{Hazay:PSI:2017}.
}
 The protocol implementation is not available, so we compare it with \sys based on the performance numbers reported by the authors, and replicate \sys's experiments on similar capacity servers.
Overall, we find that a 4-party PSI of $2^{12}$ elements per party takes \approx\SI{3}{\second} using the custom protocol in the online phase, versus \approx\SI{30}{\second} in \sys, representing a $10\times$ overhead.

\paragraph{Arithmetic MPC}
\sys builds upon a Boolean MPC framework instead of arithmetic MPC.
We validate our design choice by comparing the performance of \sys with that of SCALE-MAMBA~\cite{ScaleMamba}, a state-of-the-art arithmetic MPC framework.
We find that though arithmetic MPC is $3\times$ faster than \sys for aggregation operations \emph{alone} (as expected), this benefit doesn't generalize. 
In \sys's target workloads, aggregations are typically performed on top of operations such as joins and group by, as exemplified by our representative queries and the TPC-H query mix.
For these queries (which also represent the general case), \sys is over two orders of magnitude faster.
More specifically, we measure the latency of
\begin{enumerate*}[(i)]
\item a join with sum operation, and
\item a group by with sum operation,
\end{enumerate*} 
across 4 parties with 256 inputs per party;
we find that \sys is faster by $550\times$ and $350\times$ for the two operations, respectively.
The reason for this disparity is that joins and group by operations rely almost entirely on logical operations such as comparisons, for which Boolean MPC is much more suitable than arithmetic MPC.

\paragraph{Semi-honest systems}
\label{ss:semi_honest_eval}
We quantify the overhead of malicious security by comparing the performance of \sys with semi-honest baselines.
To the best of our knowledge, we do not know of any modern $m$-party semi-honest garbled circuit frameworks faster than AGMPC (even though it's maliciously secure).
Therefore, we implement and evaluate a semi-honest version of AGMPC ourselves, and compare \sys against it in
\cref{figure:semi_honest_performance}. %
AGMPC-SH refers to the semi-honest baseline with monolithic circuit execution.
We additionally note that \sys's techniques for decomposing circuits translate naturally to the semi-honest setting, without the need for verifying intermediate outputs.
Hence, we also implement a semi-honest version of \sys atop AGMPC-SH that decomposes queries across parties.
We do not compare \sys to prior semi-honest multi-party systems SMCQL and Conclave, as their current implementations only support $2$ to $3$ parties.

\cref{figure:semi_honest_building_blocks} plots the runtime of $\mpsi$, $\mpsu$
and $\msort$ across 16 parties, with $1K$, $600$ and $600$ inputs per party respectively.
We observe that \sys-SH yields performance benefits ranging from
$2.7$--$8.7\times$ when compared to AGMPC-SH. 
\sys's malicious security, however, comes with an overhead of $4.4\times$ compared to \sys-SH.
We also measure the end-to-end performance of the three sample queries, and find that
\sys-SH yields performance benefits
similar to \cref{figure:comorbidity_varying_input,figure:pass_varying_input,figure:cc_varying_input}
when compared to AGMPC-SH. At the same time, we observe a maximum overhead of $3.6\times$
when running the queries in a maliciously-secure setting.

\section{Limitations and Discussion}
\label{s:discussion}
\parhead{Applicability of \sys's techniques}
\sys works best for operations that can be naturally decomposed into a tree. %
While many SQL queries fit this structure, not all of them do.
A general case is one where the same relation is fed as input to two different operations (or nodes in the query tree).
For example, consider a collaboration of 3 parties, where each party $P_i$ holds a relation $R_i$, %
who wish to compute the join $(R_1 \Union R_2) \Join R_3$.
In the unencrypted setting, we can decompose the operation by computing pairwise joins 
$R_1 \Join R_3$ and $R_2 \Join R_3$, and then take the union of the results.
Unfortunately, this decomposition doesn't work in \sys because it produces a DAG (a node with two parents) and not a tree.
Hence, a malicious $P_3$ may use different values for $R_3$ across the pairwise joins, leading to an input consistency issue.
In such cases, \sys falls back to monolithic MPC for the operation.

Overall, \sys's techniques do not universally benefit all classes of computations, yet they encompass important and common analytics queries, as our sample queries exemplify.

\parheadskip{Verifiability of SQL operators}
As described in \cref{s:planning}, for simplicity, \sys's compiler requires that each node in the query tree outputs values that adhere to a well-defined set of constraints.
If a node constrains its outputs in any other way, the compiler marks it as unverifiable. 
The reason is that additional constraints restrict the space of possible inputs for future nodes in the tree (and thereby, their outputs), making it harder to deduce what needs to be verified.

For example, consider a group by operation over column $a$, with a sum over column $b$ per group.
If the values in $b$ also have a range constraint, then deducing the possible values for the sums per group is non-trivial (though technically possible).
Generalizing \sys's compiler to accept a richer (or possibly, arbitrary) set of constraints is interesting future work.

\parheadskip{Additional SQL functionality}
\sys does not support SQL operations such as UDFs, substring matching, or regular expressions, as we discuss in our analysis of the TPC-H benchmark~\cref{s:eval:tpch}.
Adding support for missing operations requires augmenting \sys's compiler to
\begin{enumerate*}[(i)]
\item translate the operation into a Boolean circuit; and
\item verify the invertibility of the operation as required by the MPC decomposition protocol.
\end{enumerate*}
While this is potentially straightforward for operations such as substring matching and (some limited types of) regular expressions, verifying the invertibility of arbitrary UDFs is computationally a hard problem.
Overall, extending \sys to support wider SQL functionality (including a well-defined class of UDFs) is an interesting direction for future work.

\parheadskip{Differential privacy}
\sys reveals the query results to all the parties, which may leak information about the underlying data samples.
This leakage can potentially be mitigated by extending \sys to support techniques such as differential privacy (DP)~\cite{Dwork:DP} (which prevents leakage by adding noise to the query results), similar to prior work~\cite{DJoin:Narayan:OSDI,Shrinkwrap:Bater:VLDB}.

In principle, one can use a general-purpose MPC protocol to implement a given DP mechanism for computing noised queries in the standard model~\cite{ODO:Dwork,Eigner14}---each party contributes a share of the randomness, which is combined within MPC to generate noise and perturb the query results, depending on the mechanism.
However, an open question is how the MPC decomposition protocol of \sys interacts with a given DP mechanism.
The mechanism governs where and how the noise is added to the computation, \eg Chorus~\cite{Chorus:Johnson} rewrites SQL queries to transform them into intrinsically private versions. On the other hand, \sys decomposes the computation across parties, which suggests that existing mechanisms may not be directly transferable to \sys in the presence of malicious adversaries while maintaining DP guarantees.
As a result, designing DP mechanisms that are compatible with \sys is a potentially interesting direction for future work.

\section{Related work}
\label{s:relatedwork}

\noindent{\bf Secure multi-party computation (MPC)} \cite{Yao:82:Millionaires,GMW87,BGW88}\textbf{.}\enskip
A variety of MPC protocols have been proposed for malicious adversaries and dishonest majority, with SPDZ \cite{Overdrive, KOS16, DKLPSS13} and WRK \cite{WRK17:Global} being the state-of-the-art for arithmetic and Boolean 
(and for multi/constant rounds) 
settings, respectively. WRK is more suited to our setting than SPDZ because relational queries map to Boolean circuits more efficiently. 
These protocols execute a given computation as a monolithic circuit.
In contrast, \sys decomposes a circuit into a tree, and executes each sub-circuit only with a subset of parties.

\parheadskip{MPC frameworks}
There are several frameworks for compiling and executing programs using MPC, in malicious~\cite{AGMPC:Git,ScaleMamba,Frigate} as well as semi-honest~\cite{ObliVM,FairplayMP,Fairplay,Picco:Zhang:2013,Sharemind,Wysteria,GraphSC} settings.
\sys builds upon the AGMPC framework~\cite{AGMPC:Git} that implements the maliciously secure WRK protocol.

\parheadskip{Private set operations}
A rich body of work exists on custom protocols for set operations (\eg \cite{Kissner:SetOps:2005,PSI:FNO18:eprint,PSU:Kolesnikov:eprint:2019,PSI:Kolesnikov:2017,PSI:PSTY:2019,SetOps:CGT:2012, PSI:CKT:2010}).
\sys's circuit primitives build upon protocols that express the set operation as a Boolean circuit~\cite{PSI:HEK12:NDSS,SetOps:Blanton:2012} in order to allow further MPC computation over the results, rather than using other primitives like oblivious transfer, oblivious PRFs, \etc.

\parheadskip{Secure collaborative systems}
Similar to \sys, recent systems such as SMCQL~\cite{SMCQL:Bater:VLDB} and Conclave~\cite{Conclave:Volgushev:Eurosys} also target privacy for collaborative query execution using MPC.
 Other proposals~\cite{ABGGKMSTX05,CLS09} support such computation by outsourcing it to two non-colluding servers.
However, all these systems assume the adversaries are semi-honest and optimize for this use case, while \sys  provides security against malicious adversaries.
Prio~\cite{Prio:Gibbs:2017}, Melis \etal~\cite{Sketches:Melis:2016}, and Prochlo~\cite{Prochlo:Bittau:2017} collect aggregate statistics across many users, as opposed to general-purpose SQL.
Further, the first two target semi-honest security, 
while Prochlo uses hardware enclaves~\cite{SGX:HASP13}.

Similar objectives have been explored for machine learning (\eg ~\cite{Federated:google,SecureAgg:google,Helen:Zheng:2019,Linear:Gascon:2017,Ridge:Nikolaenko:2013,SecureML:Mohassel,DeepML:Shokri}).
Most of these proposals target semi-honest adversaries. Others are limited to specific tasks such as linear regression, and are not applicable to \sys.

\parheadskip{Trusted hardware}
An alternate to cryptography is to use systems based on trusted hardware enclaves 
(\eg \cite{Zheng:Opaque, ObliDB, EnclaveDB}). %
Such approaches can be generalized to multi-party scenarios as well.
However, enclaves require additional trust assumptions, and suffer from many side-channel attacks~\cite{sgxattacks-foreshadow,wang-sgx-leaky}.

\parheadskip{Systems with differential privacy}
DJoin~\cite{DJoin:Narayan:OSDI} and DStress \cite{DStress:Papadimitriou} use black-box MPC protocols to compute operations over multi-party databases, and use differential privacy~\cite{Dwork:DP} to mask the results.
Shrinkwrap~\cite{Shrinkwrap:Bater:VLDB} improves the efficiency of SMCQL by using differential privacy to hide the sizes of intermediate results
(instead of padding them to an upper bound, as in \sys).
Flex~\cite{Flex:Johnson:VLDB} enforces differential privacy on the results of SQL queries, though not in the collaborative case.
In general, differential privacy solutions are complementary to \sys and can possibly be added atop \sys's processing by encoding them into \sys's circuits (as discussed in \cref{s:discussion}).

\section{Conclusion}
We presented \sys, a system for securely computing analytical SQL queries in a collaborative setup.
Unlike prior work, \sys targets a powerful adversary who may arbitrarily deviate from the specified protocol.
Compared to traditional cryptographic solutions, \sys improves performance by securely decomposing a big cryptographic computation into smaller and  parallel computations, planning an efficient decomposition, and  verifiably delegating a part of the query to local computation. 
Our techniques can improve query runtime by up to \maxperf{} when compared to the state-of-the-art.

{ 
\section*{Acknowledgments}
We thank the reviewers for their insightful feedback.
We also thank members of the RISELab at UC Berkeley for their helpful comments on earlier versions of this paper; Charles Lin for his assistance in the early phases of this project; and Carmit Hazay for valuable discussions.
This work was supported in part by the NSF CISE Expeditions Award CCF-1730628, and gifts from the Sloan Foundation, Bakar Program,  Alibaba, Amazon Web Services, Ant Group, Capital One, Ericsson, Facebook, Futurewei, Google, Intel, Microsoft, Nvidia, Scotiabank, Splunk, and VMware.
}

\let\oldthebibliography\thebibliography
\let\endoldthebibliography\endthebibliography
\renewenvironment{thebibliography}[1]{
      \begin{oldthebibliography}{#1}
              \setlength{\itemsep}{0em}
                  \setlength{\parskip}{0em}
                  
}
{
      \end{oldthebibliography}
      
}

{
\footnotesize
\begin{flushleft}
\setlength{\parskip}{0pt}
\setlength{\itemsep}{0pt}
\bibliographystyle{abbrv}
\bibliography{bib/references,bib/str_short,bib/conf}

\begin{thebibliography}{10}

\bibitem{Riskexposures:Abbe}
E.~A. Abbe, A.~E. Khandani, and A.~W. Lo.
\newblock {Privacy-Preserving Methods for Sharing Financial Risk Exposures}.
\newblock {\em American Economic Review}, 2012.

\bibitem{AfsharHMR15}
A.~Afshar, Z.~Hu, P.~Mohassel, and M.~Rosulek.
\newblock How to efficiently evaluate {RAM} programs with malicious security.
\newblock In {\em EUROCRYPT}, 2015.

\bibitem{ABGGKMSTX05}
G.~Aggarwal, M.~Bawa, P.~Ganesan, H.~Garcia{-}Molina, K.~Kenthapadi,
  R.~Motwani, U.~Srivastava, D.~Thomas, and Y.~Xu.
\newblock Two can keep {A} secret: {A} distributed architecture for secure
  database services.
\newblock In {\em CIDR}, 2005.

\bibitem{SMCQL:Bater:VLDB}
J.~Bater, G.~Elliott, C.~Eggen, S.~Goel, A.~Kho, and J.~Rogers.
\newblock {SMCQL: Secure Querying for Federated Databases}.
\newblock In {\em VLDB}, 2017.

\bibitem{Shrinkwrap:Bater:VLDB}
J.~Bater, X.~He, W.~Ehrich, A.~Machanavajjhala, and J.~Rogers.
\newblock {Shrinkwrap: Differentially-Private Query Processing in Private Data
  Federations}.
\newblock In {\em VLDB}, 2018.

\bibitem{BMR90}
D.~Beaver, S.~Micali, and P.~Rogaway.
\newblock The round complexity of secure protocols (extended abstract).
\newblock In {\em STOC}, 1990.

\bibitem{BHR12}
M.~Bellare, V.~T. Hoang, and P.~Rogaway.
\newblock Foundations of garbled circuits.
\newblock In {\em CCS}, 2012.

\bibitem{FairplayMP}
A.~Ben-David, N.~Nisan, and B.~Pinkas.
\newblock {FairplayMP: A System for Secure Multi-party Computation}.
\newblock In {\em CCS}, 2008.

\bibitem{BGW88}
M.~Ben{-}Or, S.~Goldwasser, and A.~Wigderson.
\newblock Completeness theorems for non-cryptographic fault-tolerant
  distributed computation (extended abstract).
\newblock In {\em STOC}, 1988.

\bibitem{Riskanalytics:Bisias}
D.~Bisias, M.~Flood, A.~W. Lo, and S.~Valavanis.
\newblock {A Survey of Systemic Risk Analytics}.
\newblock {\em Annual Review of Financial Economics}, 2012.

\bibitem{Prochlo:Bittau:2017}
A.~Bittau et~al.
\newblock {Prochlo: Strong Privacy for Analytics in the Crowd}.
\newblock In {\em SOSP}, 2017.

\bibitem{SetOps:Blanton:2012}
M.~Blanton and E.~Aguiar.
\newblock {Private and oblivious set and multiset operations}.
\newblock In {\em AsiaCCS}, 2012.

\bibitem{arxrange}
T.~Boelter, R.~Poddar, and R.~A. Popa.
\newblock {A Secure One-Roundtrip Index for Range Queries}.
\newblock Cryptology ePrint Archive, Report 2016/568, 2016.
\newblock \url{https://eprint.iacr.org/2016/568}.

\bibitem{Sharemind}
D.~Bogdanov, S.~Laur, and J.~Willemson.
\newblock {Sharemind: A framework for fast privacy-preserving computations}.
\newblock In {\em ESORICS}, 2008.

\bibitem{SecureAgg:google}
K.~Bonawitz et~al.
\newblock {Practical Secure Aggregation for Privacy-Preserving Machine
  Learning}.
\newblock In {\em CCS}, 2017.

\bibitem{sgxattacks-foreshadow}
J.~V. Bulck et~al.
\newblock {Foreshadow: Extracting the Keys to the Intel {SGX} Kingdom with
  Transient Out-of-Order Execution}.
\newblock In {\em USENIX Security}, 2018.

\bibitem{cdcdiseases}
{Center for Disease Control and Prevention (CDC): Diseases and Conditions A-Z
  Index}, 2017.
\newblock \url{https://www.cdc.gov/DiseasesConditions}.

\bibitem{CJS12}
J.~H. Cheon, S.~Jarecki, and J.~H. Seo.
\newblock Multi-party privacy-preserving set intersection with quasi-linear
  complexity.
\newblock {\em {IEICE} Transactions}, 95-A(8):1366--1378, 2012.

\bibitem{CLS09}
S.~S.~M. Chow, J.~Lee, and L.~Subramanian.
\newblock Two-party computation model for privacy-preserving queries over
  distributed databases.
\newblock In {\em NDSS}, 2009.

\bibitem{Codd:relational}
E.~F. Codd.
\newblock {A Relational Model of Data for Large Shared Data Banks}.
\newblock {\em Commun. ACM}, 1970.

\bibitem{Prio:Gibbs:2017}
H.~Corrigan-Gibbs and D.~Boneh.
\newblock {Prio: Private, Robust, and Scalable Computation of Aggregate
  Statistics}.
\newblock In {\em NSDI}, 2017.

\bibitem{SetOps:CGT:2012}
E.~D. Cristofaro, P.~Gasti, and G.~Tsudik.
\newblock {Fast and Private Computation of Cardinality of Set Intersection and
  Union}.
\newblock In {\em CANS}, 2012.

\bibitem{PSI:CKT:2010}
E.~D. Cristofaro, J.~Kim, and G.~Tsudik.
\newblock {Linear-Complexity Private Set Intersection Protocols Secure in
  Malicious Model}.
\newblock In {\em ASIACRYPT}, 2010.

\bibitem{DMRY11}
D.~Dachman-Soled, T.~Malkin, M.~Raykova, and M.~Yung.
\newblock {Secure Efficient Multiparty Computing of Multivariate Polynomials
  and Applications}.
\newblock In {\em ACNS}, 2011.

\bibitem{DKLPSS13}
I.~Damg{\aa}rd, M.~Keller, E.~Larraia, V.~Pastro, P.~Scholl, and N.~P. Smart.
\newblock Practical covertly secure {MPC} for dishonest majority - or: Breaking
  the {SPDZ} limits.
\newblock In {\em ESORICS}, 2013.

\bibitem{DIRTYSOCK:UBUNTU}
Privilege {E}scalation in {U}buntu {L}inux, 2019.
\newblock \url{https://shenaniganslabs.io/2019/02/13/Dirty-Sock.html}.

\bibitem{ODO:Dwork}
C.~Dwork, K.~Kenthapadi, F.~McSherry, I.~Mironov, and M.~Naor.
\newblock {Our Data, Ourselves: Privacy via Distributed Noise Generation}.
\newblock In {\em EUROCRYPT}, 2006.

\bibitem{Dwork:DP}
C.~Dwork and A.~Roth.
\newblock {The Algorithmic Foundations of Differential Privacy}.
\newblock {\em Found. Trends Theor. Comput. Sci.}, 2014.

\bibitem{Eigner14}
F.~Eigner, A.~Kate, M.~Maffei, F.~Pampaloni, and I.~Pryvalov.
\newblock Differentially private data aggregation with optimal utility.
\newblock In {\em ACSAC}, 2014.

\bibitem{AGMPC:Git}
{AGMPC Framework}.
\newblock \url{https://github.com/emp-toolkit/emp-agmpc}.

\bibitem{ObliDB}
S.~Eskandarian and M.~Zaharia.
\newblock {ObliDB: Oblivious Query Processing using Hardware Enclaves}.
\newblock 2020.

\bibitem{PSI:FNO18:eprint}
B.~H. Falk, D.~Noble, and R.~Ostrovsky.
\newblock {Private Set Intersection with Linear Communication from General
  Assumptions}.
\newblock In {\em WPES}, 2019.

\bibitem{FrederiksenJNNO13}
T.~K. Frederiksen, T.~P. Jakobsen, J.~B. Nielsen, P.~S. Nordholt, and
  C.~Orlandi.
\newblock {MiniLEGO: Efficient Secure Two-Party Computation from General
  Assumptions}.
\newblock In {\em EUROCRYPT}, 2013.

\bibitem{FrederiksenJNT15}
T.~K. Frederiksen, T.~P. Jakobsen, J.~B. Nielsen, and R.~Trifiletti.
\newblock {TinyLEGO: An Interactive Garbling Scheme for Maliciously Secure
  Two-party Computation}.
\newblock Cryptology ePrint Archive, Report 2015/309, 2015.
\newblock \url{https://eprint.iacr.org/2015/309}.

\bibitem{Garg0MP16}
S.~Garg, D.~Gupta, P.~Miao, and O.~Pandey.
\newblock Secure multiparty {RAM} computation in constant rounds.
\newblock In {\em TCC}, 2016.

\bibitem{GargLOS15}
S.~Garg, S.~Lu, R.~Ostrovsky, and A.~Scafuro.
\newblock Garbled {RAM} from one-way functions.
\newblock In {\em STOC}, 2015.

\bibitem{Linear:Gascon:2017}
A.~Gasc\'{o}n, P.~Schoppmann, B.~Balle, M.~Raykova, J.~Doerner, S.~Zahur, and
  D.~Evans.
\newblock {Privacy-Preserving Distributed Linear Regression on High-Dimensional
  Data}.
\newblock In {\em PETS}, 2017.

\bibitem{Goldreich2004}
O.~Goldreich.
\newblock {\em The Foundations of Cryptography - Volume 2: Basic Applications}.
\newblock Cambridge University Press, 2004.

\bibitem{GMW87}
O.~Goldreich, S.~Micali, and A.~Wigderson.
\newblock {How to Play ANY Mental Game}.
\newblock In {\em STOC}, 1987.

\bibitem{Federated:google}
{Google AI}.
\newblock {Federated Learning: Collaborative Machine Learning without
  Centralized Training Data}.
\newblock
  \url{https://ai.googleblog.com/2017/04/federated-learning-collaborative.html}.

\bibitem{Hazay:PSI:2017}
C.~Hazay and M.~Venkitasubramaniam.
\newblock {Scalable Multi-Party Private Set-Intersection}.
\newblock In {\em PKC}, 2017.

\bibitem{HazayY16}
C.~Hazay and A.~Yanai.
\newblock Constant-round maliciously secure two-party computation in the {RAM}
  model.
\newblock In {\em TCC}, 2016.

\bibitem{PSI:HEK12:NDSS}
Y.~Huang, D.~Evans, and J.~Katz.
\newblock {Private Set Intersection: Are Garbled Circuits Better than Custom
  Protocols?}
\newblock In {\em NDSS}, 2012.

\bibitem{PSIsum:Google:2017}
M.~Ion et~al.
\newblock {Private Intersection-Sum Protocol with Applications to Attributing
  Aggregate Ad Conversions}.
\newblock Cryptology ePrint Archive, Report 2017/738, 2017.
\newblock \url{https://eprint.iacr.org/2017/738}.

\bibitem{Flex:Johnson:VLDB}
N.~Johnson, J.~P. Near, and D.~Song.
\newblock {Towards Practical Differential Privacy for SQL Queries}.
\newblock In {\em VLDB}, 2018.

\bibitem{Chorus:Johnson}
N.~M. Johnson, J.~P. Near, J.~M. Hellerstein, and D.~Song.
\newblock {Chorus: Differential Privacy via Query Rewriting}.
\newblock {\em arXiv:1809.07750}, 2018.

\bibitem{GenomeAssociation:Kamm}
L.~Kamm, D.~Bogdanov, and J.~Vilo.
\newblock {A new way to protect privacy in large-scale genome-wide association
  studies}.
\newblock {\em Bioinformatics}, 2013.

\bibitem{KOS16}
M.~Keller, E.~Orsini, and P.~Scholl.
\newblock {MASCOT:} faster malicious arithmetic secure computation with
  oblivious transfer.
\newblock In {\em CCS}, 2016.

\bibitem{Overdrive}
M.~Keller, V.~Pastro, and D.~Rotaru.
\newblock {Overdrive: Making SPDZ Great Again}.
\newblock In {\em EUROCRYPT}, 2018.

\bibitem{KellerY18}
M.~Keller and A.~Yanai.
\newblock Efficient maliciously secure multiparty computation for {RAM}.
\newblock In {\em EUROCRYPT}, 2018.

\bibitem{Kissner:SetOps:2005}
L.~Kissner and D.~Song.
\newblock {Privacy-Preserving Set Operations}.
\newblock In {\em CRYPTO}, 2005.

\bibitem{PSI:Kolesnikov:2017}
V.~Kolesnikov, N.~Matania, B.~Pinkas, M.~Rosulek, and N.~Trieu.
\newblock {Practical Multi-party Private Set Intersection from Symmetric-Key
  Techniques}.
\newblock In {\em CCS}, 2017.

\bibitem{KolesnikovNRTT17}
V.~Kolesnikov, J.~B. Nielsen, M.~Rosulek, N.~Trieu, and R.~Trifiletti.
\newblock {DUPLO:} unifying cut-and-choose for garbled circuits.
\newblock In {\em CCS}, 2017.

\bibitem{PSU:Kolesnikov:eprint:2019}
V.~Kolesnikov, M.~Rosulek, N.~Trieu, and X.~Wang.
\newblock {Scalable Private Set Union from Symmetric-Key Techniques}.
\newblock In {\em ASIACRYPT}, 2019.

\bibitem{ObliVM}
C.~Liu, X.~S. Wang, K.~Nayak, Y.~Huang, and E.~Shi.
\newblock {ObliVM: A Programming Framework for Secure Computation}.
\newblock In {\em IEEE S\&P}, 2015.

\bibitem{LuO17}
S.~Lu and R.~Ostrovsky.
\newblock Black-box parallel garbled {RAM}.
\newblock In {\em CRYPTO}, 2017.

\bibitem{Fairplay}
D.~Malkhi, N.~Nisan, B.~Pinkas, and Y.~Sella.
\newblock {Fairplay -- A Secure Two-party Computation System}.
\newblock In {\em USENIX Security}, 2004.

\bibitem{SGX:HASP13}
F.~McKeen et~al.
\newblock {Innovative Instructions and Software Model for Isolated Execution}.
\newblock In {\em HASP}, 2013.

\bibitem{Sketches:Melis:2016}
L.~Melis, G.~Danezis, and E.~D. Cristofaro.
\newblock {Efficient Private Statistics with Succinct Sketches}.
\newblock In {\em NDSS}, 2016.

\bibitem{SecureML:Mohassel}
P.~Mohassel and Y.~Zhang.
\newblock {SecureML: A System for Scalable Privacy-Preserving Machine
  Learning}.
\newblock In {\em IEEE S\&P}, 2019.

\bibitem{Frigate}
B.~Mood, D.~Gupta, H.~Carter, K.~R.~B. Butler, and P.~Traynor.
\newblock {Frigate: A Validated, Extensible, and Efficient Compiler and
  Interpreter}.
\newblock In {\em EuroS\&P}, 2016.

\bibitem{DJoin:Narayan:OSDI}
A.~Narayan and A.~Haeberlen.
\newblock {DJoin: Differentially Private Join Queries over Distributed
  Databases}.
\newblock In {\em OSDI}, 2012.

\bibitem{GraphSC}
K.~Nayak, X.~S. Wang, S.~Ioannidis, U.~Weinsberg, N.~Taft, and E.~Shi.
\newblock {GraphSC: Parallel Secure Computation Made Easy}.
\newblock In {\em IEEE S\&P}, 2015.

\bibitem{NNOB12:TinyOT}
J.~B. Nielsen, P.~S. Nordholt, C.~Orlandi, and S.~S. Burra.
\newblock {A New Approach to Practical Active-Secure Two-Party Computation}.
\newblock In {\em CRYPTO}, 2012.

\bibitem{NielsenO09}
J.~B. Nielsen and C.~Orlandi.
\newblock {LEGO} for two-party secure computation.
\newblock In {\em TCC}, 2009.

\bibitem{Ridge:Nikolaenko:2013}
V.~Nikolaenko, U.~Weinsberg, S.~Ioannidis, M.~Joye, D.~Boneh, and N.~Taft.
\newblock {Privacy-Preserving Ridge Regression on Hundreds of Millions of
  Records}.
\newblock In {\em IEEE S\&P}, 2013.

\bibitem{DStress:Papadimitriou}
A.~Papadimitriou, A.~Narayan, and A.~Haeberlen.
\newblock {DStress: Efficient Differentially Private Computations on
  Distributed Data}.
\newblock In {\em EuroSys}, 2017.

\bibitem{SHELLSHOCK:NYT}
N.~Perlroth.
\newblock Security {E}xperts {E}xpect `{S}hellshock' {S}oftware {B}ug in {B}ash
  to {B}e {S}ignificant, 2014.
\newblock
  \url{https://www.nytimes.com/2014/09/26/technology/security-experts-expect-shellshock-software-bug-to-be-significant.html}.

\bibitem{PSI:PSTY:2019}
B.~Pinkas, T.~Schneider, O.~Tkachenko, and A.~Yanai.
\newblock {Efficient Circuit-Based {PSI} with Linear Communication}.
\newblock In {\em EUROCRYPT}, 2019.

\bibitem{arx}
R.~Poddar, T.~Boelter, and R.~A. Popa.
\newblock {Arx: An Encrypted Database using Semantically Secure Encryption}.
\newblock In {\em VLDB}, 2019.

\bibitem{EnclaveDB}
C.~Priebe, K.~Vasawani, and M.~Costa.
\newblock {EnclaveDB: A Secure Database Using SGX}.
\newblock In {\em IEEE S\&P}, 2018.

\bibitem{Wysteria}
A.~Rastogi, M.~A. Hammer, and M.~Hicks.
\newblock {Wysteria: A Programming Language for Generic, Mixed-Mode Multiparty
  Computations}.
\newblock In {\em IEEE S\&P}, 2014.

\bibitem{Frauddetection:Sangers:2018}
A.~Sangers, M.~van Heesch, T.~Attema, T.~Veugen, M.~Wiggerman, J.~Veldsink,
  O.~Bloemen, and D.~Worm.
\newblock {Secure multiparty PageRank algorithm for collaborative fraud
  detection}.
\newblock In {\em FC}, 2019.

\bibitem{ScaleMamba}
{SCALE-MAMBA Framework}.
\newblock \url{https://homes.esat.kuleuven.be/~nsmart/SCALE/}.

\bibitem{DeepML:Shokri}
R.~Shokri and V.~Shmatikov.
\newblock {Privacy-Preserving Deep Learning}.
\newblock In {\em CCS}, 2015.

\bibitem{TPCH:Web}
{T{PC}-{H} Benchmark}.
\newblock \url{http://www.tpc.org/tpch/}.

\bibitem{Conclave:Volgushev:Eurosys}
N.~Volgushev, M.~Schwarzkopf, B.~Getchell, M.~Varia, A.~Lapets, and
  A.~Bestavros.
\newblock {Conclave: Secure Multi-Party Computation on Big Data}.
\newblock In {\em EuroSys}, 2019.

\bibitem{PASSREUSE:WR19:NDSS}
K.~C. Wang and M.~K. Reiter.
\newblock How to end password reuse on the web.
\newblock In {\em NDSS}, 2019.

\bibitem{wang-sgx-leaky}
W.~Wang, G.~Chen, X.~Pan, Y.~Zhang, X.~Wang, V.~Bindschaedler, H.~Tang, and
  C.~A. Gunter.
\newblock {Leaky Cauldron on the Dark Land: Understanding Memory Side-Channel
  Hazards in SGX}.
\newblock In {\em CCS}, 2017.

\bibitem{WRK17:Global}
X.~Wang, S.~Ranellucci, and J.~Katz.
\newblock {Global-Scale Secure Multiparty Computation}.
\newblock In {\em CCS}, 2017.

\bibitem{Yao:82:Millionaires}
A.~C. Yao.
\newblock Protocols for secure computations.
\newblock In {\em Symposium on Foundations of Computer Science (SFCS)}, 1982.

\bibitem{Yao86:GC}
A.~C. Yao.
\newblock {How to generate and exchange secrets (extended abstract)}.
\newblock In {\em FOCS}, 1986.

\bibitem{Zhang:CCSW}
E.~Zhang, F.-H. Liu, Q.~Lai, G.~Jin, and Y.~Li.
\newblock {Efficient Multi-Party Private Set Intersection Against Malicious
  Adversaries}.
\newblock In {\em CCSW}, 2019.

\bibitem{Picco:Zhang:2013}
Y.~Zhang, A.~Steele, and M.~Blanton.
\newblock {PICCO: A General-purpose Compiler for Private Distributed
  Computation}.
\newblock In {\em CCS}, 2013.

\bibitem{Zheng:Opaque}
W.~Zheng, A.~Dave, J.~G. Beekman, R.~A. Popa, J.~E. Gonzalez, and I.~Stoica.
\newblock {Opaque: An Oblivious and Encrypted Distributed Analytics Platform}.
\newblock In {\em NSDI}, 2017.

\bibitem{Helen:Zheng:2019}
W.~Zheng, R.~A. Popa, J.~Gonzalez, and I.~Stoica.
\newblock {Helen: Maliciously Secure Coopetitive Learning for Linear Models}.
\newblock In {\em IEEE S\&P}, 2019.

\end{thebibliography}
\end{flushleft}
}

\end{document}